\documentclass[lettersize,journal]{IEEEtran}
\usepackage{amsmath,amsfonts}

\usepackage{algorithm}
\usepackage{algorithmicx}
\usepackage{algpseudocode}
\usepackage[inline]{enumitem}

\usepackage{array}
\usepackage[caption=false,font=normalsize,labelfont=sf,textfont=sf]{subfig}
\usepackage{amssymb}
\usepackage{textcomp}
\usepackage{stfloats}
\usepackage{placeins} 
\usepackage{url}
\usepackage{verbatim}
\usepackage{graphicx}
\usepackage{xcolor}
\usepackage{xcolor,colortbl}
\usepackage[
  separate-uncertainty = true,
  multi-part-units = repeat,
  parse-numbers=false
]{siunitx}
\usepackage{multirow}
\usepackage{booktabs}

\usepackage[style=ieee, hyperref=false]{biblatex}
\AtEveryBibitem{
    \clearfield{url}
    \clearfield{urlyear}
    \clearfield{urlmonth}
    \clearfield{doi}
    \clearfield{issn}
    \clearfield{eprint}
}

\AtBeginBibliography{\footnotesize}

\algblock{ParFor}{EndParFor}
\algnewcommand\algorithmicparfor{\textbf{for}}
\algnewcommand\algorithmicpardo{\textbf{do in parallel}}
\algnewcommand\algorithmicendparfor{\textbf{end}}
\algrenewtext{ParFor}[1]{\algorithmicparfor\ #1\ \algorithmicpardo}
\algrenewtext{EndParFor}{\algorithmicendparfor}

\begin{document}

\title{Parallel Branch Model Predictive Control on GPUs}

\author{Luyao Zhang, Chenghuai Lin, Sergio Grammatico
\thanks{*This work is partially supported by NWO under project AMADeuS.}
\thanks{Luyao Zhang, Chenghuai Lin, and Sergio Grammatico are with the Delft Center for Systems and Control, TU Delft, The Netherlands.
        {\tt\small \{l.zhang-7, chenghuailin,
        s.grammatico\}@tudelft.nl}.
        }%
}

\markboth{Journal of \LaTeX\ Class Files,~Vol.~14, No.~8, August~2021}%
{Shell \MakeLowercase{\textit{et al.}}: A Sample Article Using IEEEtran.cls for IEEE Journals}


\maketitle

\begin{abstract}
We present a GPU-based solver for trajectory planning problems using branch model predictive control.
Building on iterative LQR methods, we adopt a multiple-shooting formulation for the system dynamics and use an augmented Lagrangian method to handle general stage-wise constraints. 
This design enables straightforward warm-starting.
The constraint-handling capability of our solver is validated on two challenging trajectory planning problems.
In addition, we develop two tailored inner LQR solvers that exploit the tree-sparse structure.
The solvers offer different levels of parallelism, making them appropriate for different tree sizes.
The numerical results demonstrate that, compared to a high-performance CPU-based solver, our approach achieves superior performance on large-scale problems.

\end{abstract}


\section{Introduction}
Model predictive control (MPC) is a state-of-the-art planning method that numerically solves optimal control problems in a receding horizon fashion.
MPC has gained popularity in practice due to its ability to handle constraints and adapt to dynamic environments.
In recent years, this approach has achieved significant success in its application to mobile robots \cite{WensingOptimization2024, ChenCiLQR2019, GrandiaLocomotion2023, NguyenAerial2021}.
Nominal MPC does not explicitly account for uncertainty, often resulting in undesirable planning performance. 
For example, when an automated vehicle attempts an unprotected left turn, the surrounding vehicle needs to decide whether to yield to it or not.
If the automated vehicle misinterprets the surrounding vehicle's driving intention, it may need to make abrupt braking maneuvers to ensure safety, resulting in an uncomfortable ride.

Branch model predictive control (BMPC)~\cite{ChenBranchMPC2022} models uncertainty using a discrete distribution. Related formulations include contingency MPC~\cite{HardyContingencyPlanning2013, AlsterdaContingency2019}, scenario tree MPC~\cite{BatkovicRobustSMPC2021}, and multistage MPC~\cite{FRISON201714399, kouzoupis_dual_2019}.
In the context of automated driving, an advantage of BMPC planners is their ability to leverage multi-modal predictions generated by motion prediction modules. 
Specifically, BMPC plans a trajectory tree with multiple branches, each corresponding to a different uncertainty realization. 
In contrast, traditional robust motion planners \cite{WangNonGaussian2020, AhnSafeMotionPlanning2022} attempt to generate a single motion plan that accommodates all predicted scenarios. 
Consequently, BMPC can avoid overly conservative motion plans.
Furthermore, recent research has successfully integrated large language models with BMPC, enhancing the trustworthiness of the system~\cite{sinha_real-time_2024}.

While BMPC demonstrates considerable promise for modeling uncertainty in real-world applications, its widespread deployment is hindered by the significant computational time required to solve the BMPC problem with long planning horizons and numerous predicted scenarios. 
Therefore, it is necessary to develop customized BMPC solvers that fully exploit the inherent tree structure. 
A common approach is to decompose the trajectory tree into distinct scenarios that can be processed in parallel on multi-core CPUs~\cite{PhiquepalControlTree2021} and GPUs~\cite{sampathirao_massively_2024}. 
The coupling between scenarios is established by enforcing the nonanticipativity constraints, ensuring that the control input does not anticipate the realization of uncertainty.
However, the use of Lagrangian relaxation in these methods increases the number of iterations required for convergence, making them unsuitable for real-time applications~\cite{marti_improving_2015}.
To respect nonanticipativity constraints exactly, several factorization techniques that exploit tree structure have been proposed~\cite{marti_improving_2015, kouzoupis_dual_2019, FRISON201714399}.
Our method falls into this category by performing a modified Riccati recursion directly on the tree.

While the aforementioned methods carefully consider the inherent tree structure in BMPC, they fail to achieve temporal parallelism.
This type of parallelism is well-suited for applications requiring a long planning horizon, such as autonomous parking \cite{HeTDROBCA2021} and planning for low-thrust spacecraft \cite{BettsLowThrust2003}.
Several parallel approaches have been proposed for the linear-quadratic (LQ) problem.
One category follows the idea of ``divide and conquer'', breaking the problem into smaller subproblems that can be solved independently before combining the results.
In \cite{Plancher_performance_2020, FarshidianMPC2017, DengParallel2019}, the subproblem can only be solved approximately because it relies on information from the previous iteration.
This methodology may fail if the trajectories between iterations diverge significantly. To address this limitation, some research efforts have focused on developing exact parallel algorithms.
Wright \cite{wright_partitioned_1991} introduces two parallel algorithms, PDP and PRI, which differ in how they solve the linear systems associated with the LQ problem and how they parameterize the subproblems.
Nielsen and Axehill \cite{nielsen_parallel_2015} treat the terminal cost in the subproblem as the unknown part and introduce an additional control input term to compensate for its influence.
This approach shares a concept similar to that of PDP, but with the difference that the consensus problem here remains an optimal control problem.
Compared with the approach in~\cite{wright_partitioned_1991}, Laine and Tomlin \cite{laine_parallelizing_2019} parameterize the optimal state and input trajectories, as well as the multipliers for the subproblem (fixed-end LQR), using only the initial and terminal states. 
A recent work \cite{SarkkaTemporal2023}, closely related to PRI, interprets the LQ problem as a cumulative sum problem and takes advantage of the parallel scan algorithm \cite{Blelloch1990PrefixSA} to solve it efficiently on GPUs.
In \cite{jallet_parallel_2024}, similarly to \cite{SarkkaTemporal2023}, the authors formulate the consensus problem as a min-max problem. They approach the subproblem through parametric optimization and perform a generalized Riccati recursion to derive the parameters for the consensus problem.

Building on prior work, we introduce a parallel BMPC solver that supports nonlinear dynamics and general stage-wise constraints.
Unlike \cite{PhiquepalControlTree2021}, our approach satisfies the nonanticipativity constraints induced by the tree structure exactly.
Our contributions are summarized as follows:
\begin{enumerate*}
    \item Inspired by the LQR solver in \cite{SarkkaTemporal2023}, we develop two parallel tree LQR solvers that achieve different levels of parallelism, as summarized in Table~\ref{tab:summary}, allowing users to choose based on the problem size and available computational resources.
    \item Building upon the proposed tree LQR solvers, we introduce a multiple-shooting iterative tree LQR solver for nonlinear BMPC problems and incorporate constraint handling using an augmented Lagrangian method.
    \item We benchmark our proposed solver against other iLQR solvers, including \textsc{Trajax} \cite{trajax} and \textsc{MPX} \cite{AmatucciMPX2026}, and release the open-source implementation.  
\end{enumerate*}

The paper is structured as follows. In Section \ref{sec:background}, we review the parallel scan and the BMPC formulation. 
Section \ref{sec:parallel_tree_lqr} discusses the parallel method for solving tree LQR problems.
In Section \ref{sec:parallel_bmpc}, we introduce an augmented Lagrangian-based method for addressing constrained nonlinear BMPC problems.
We then evaluate the computational performance of the proposed method in Section \ref{sec:results}.
Section \ref{sec:conclusion} concludes the paper.  

\begin{table}[t]
    \caption{Comparisons of different tree MPC solvers. NT: Nonlinear Tree. LT: Linear Tree. NC: Nonanticipativity Constraint. SC: scenario}
    \centering
    \renewcommand{\arraystretch}{1.2}
    \setlength{\tabcolsep}{3pt} 
    \begin{tabular}{@{} c c c c c @{}}
        \toprule
        \textbf{Solver} & \textbf{Problem Type} & \textbf{NC handling}  & \textbf{Parallelization} & \textbf{Devices} \\
        \midrule
        \textsc{sLQR} (Ours) & NT MPC  & Exact & SC & CPU/GPU \\
        \textsc{stLQR} (Ours) & NT MPC  & Exact & SC + Time & CPU/GPU \\
        \midrule
        \textsc{CT}~\cite{PhiquepalControlTree2021} & NT MPC & Inexact & SC & CPU \\
        \midrule
        \textsc{HPIPM}~\cite{FRISON201714399} & LT MPC & Exact & $\times$ & CPU \\
        \midrule
        \textsc{treeQP}~\cite{kouzoupis_dual_2019} & LT MPC & Exact & SC & CPU \\
        \midrule
        \textsc{NAMA}~\cite{sampathirao_massively_2024} & LT MPC  & Exact & SC & GPU \\
        \bottomrule
    \end{tabular}
    \label{tab:summary}
\end{table}

\section{Mathematical Background} \label{sec:background}

\begin{figure}
    \centering
    \includegraphics[width=0.90\columnwidth]{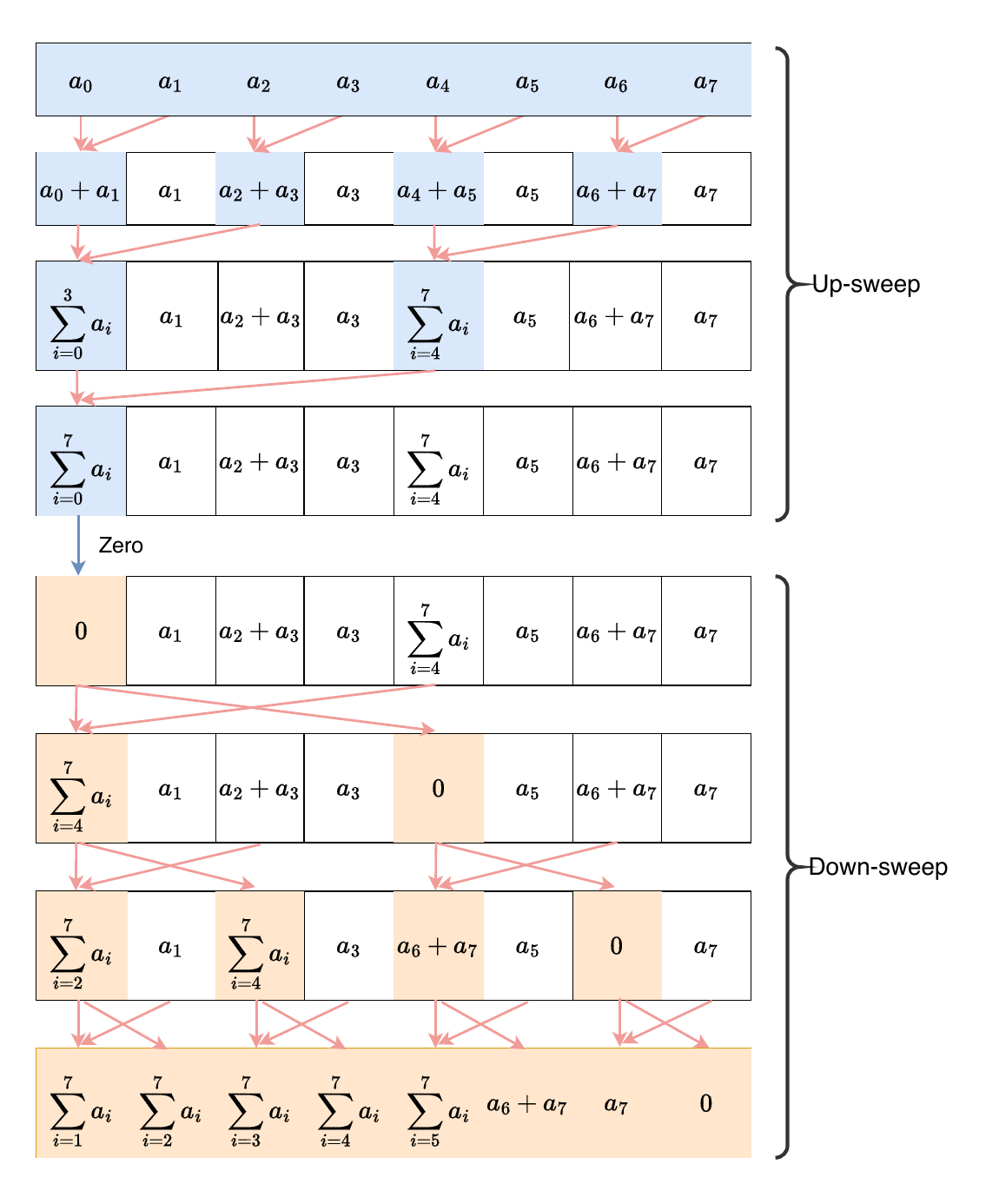}
    \caption{Parallel scan applied to the reverse cumulative sum problem. We process
    the array using a balanced binary tree. The tree nodes are highlighted in color.
    A reduction operation is performed in the up-sweep phase, leading to $\sum_{i=0}^{7} a_i$.
    Then, a down-sweep is conducted, starting from the root of the computation tree.
    For instance, at the first level, the value of the root is passed to its right and left children, and the value of the root's right child in the up-sweep $\sum_{i=4}^{7} a_i$ is added to the root's left child in the down-sweep.}
    \label{fig:parallel_scan}
\end{figure}

\subsection{Parallel Scan for LQR Problems}
The LQR problem serves as a fundamental building block for nonlinear optimal control problems and is formulated as
\begin{equation} \label{eq:LQR}
    \begin{split} 
    \min _{\boldsymbol{x}, \boldsymbol{u}} &\sum _{k=0}^{N-1}\begin{bmatrix}
        q_{k} \\  r_{k} 
    \end{bmatrix} ^\top 
    \begin{bmatrix} x_k \\ u_k \end{bmatrix} + \frac{1}{2}
    \begin{bmatrix} x_k \\ u_k \end{bmatrix} ^\top 
    \begin{bmatrix}
        Q_{k} &  M_{k}^\top\\ M_{k} & R_{k}
    \end{bmatrix} 
    \begin{bmatrix} x_k \\ u_k \end{bmatrix} \\ 
    &\quad\quad + q_N^\top x_N + \frac{1}{2} x_N^\top Q_N x_N \\
    \text{s.t.} &\; x_{k+1} = A_kx_k + B_ku_k + c_k,\; \forall k \in [0, N-1],
    \end{split} 
\end{equation}
where $\boldsymbol{x} := (x_k)_{k \in [1, N]}$, $\boldsymbol{u} := (u_k)_{k \in [0, N-1]}$, and $N$ is the horizon length. Throughout, $x_0$ is a fixed parameter.
This formulation reveals an interesting connection to parallel scan methods~\cite{Blelloch1990PrefixSA}. 
Therefore, let us briefly review the parallel scan in the context of LQR problems~\cite{SarkkaTemporal2023}.
We start by considering a problem where the input is an array of $N$ elements, 
\begin{equation}
 [a_0, a_1, \dots, a_{N-1}]
 \label{eq:array_element},
\end{equation}
and the objective is to output the following array
\begin{equation}
    [a_0, (a_0 \oplus a_1), \dots, (a_0 \oplus a_1 \oplus \dots \oplus a_{N-1})],
    \label{eq:output_array_element}
\end{equation}
where each element is the ``cumulative sum'' of the previous elements in the input array.
The symbol $\oplus$ denotes an associative operator, such as addition or multiplication. 
We also consider a similar output array, but in reverse order: $[(a_0 \oplus a_1 \oplus \dots \oplus a_{N-1}), \dots, (a_{N-2} \oplus a_{N-1}), a_{N-1}]$.
This ``cumulative sum'' problem can be solved with a time complexity of $O(\log N)$ using the parallel scan.
Figure \ref{fig:parallel_scan} illustrates the parallel scan applied to an array of 8 elements.
We process the array using a balanced binary tree. 
During the up-sweep phase, a reduction operation is performed, resulting in the value $a_0 \oplus a_1 \oplus \dots \oplus a_{N-1}$.
Next, a down-sweep is executed, starting from the root and moving toward the leaves. At each level of the tree, the value of each node is copied to its right child, while its value, updated by adding the right child's value in the up-sweep phase, is sent to the left child.
To establish its connection with the LQR problem, we define the conditional value function (CVF) \cite{SarkkaTemporal2023}, $V_{k\to i}: \mathbb{R}^{n_x}\times \mathbb{R}^{n_x} \to \mathbb{R}$, which represents the cost of the optimal trajectory starting from $x_k$ and ending at $x_i$.
The composition of two consecutive CVFs is obtained by minimizing their sum over the intermediate state:
\begin{align} \label{eq:V_ki_min}
    V_{k\to i}(x_k, x_i) &= V_{k\to j}(x_k, x_j) \oplus V_{j\to i}(x_j, x_i) \\
    &= \min_{x_j}\; V_{k\to j}(x_k, x_j) + V_{j\to i}(x_j, x_i), \nonumber 
\end{align}
for $k < j < i \leq N$.
From \cite{SarkkaTemporal2023}, $V_{k\to i}$ is given by 
\begin{equation} \label{eq:sol_min}
\begin{aligned} 
    V_{k\to i}(x_k, x_i)
    = &\max_{\lambda}\; z +\frac{1}{2}x_{k}^{\top}P_{k,i}x_{k} + p_{k,i}^{\top} x_{k} \\ &-\frac{1}{2}\lambda ^{\top }C_{k,i}\lambda {+}\lambda ^{\top }\left(x_{i}{-}A_{k,i}x_{k}{-}c_{k,i}\right),
\end{aligned}
\end{equation}
where $z$ is a constant, and $\lambda \in \mathbb{R}^{n_x}$ represents the Lagrange multiplier associated with the terminal state constraint. 
The matrices in~\eqref{eq:sol_min} are derived analytically by addressing the minimization problem in~\eqref{eq:V_ki_min} and are given by
\begin{equation} 
\begin{split} \label{eq:comb_rule}
A_{k,i} &= A_{j,i} (I + C_{k,j} P_{j,i})^{-1} A_{k,j}, \\ 
c_{k,i} &= A_{j,i} (I + C_{k,j} P_{j,i})^{-1} (c_{k,j} - C_{k,j} p_{j,i}) + c_{j,i}, \\ 
C_{k,i} &= A_{j,i} (I + C_{k,j} P_{j,i})^{-1} C_{k,j} A_{j,i}^\top + C_{j,i}, \\ 
p_{k,i} &= A_{k,j}^\top (I + P_{j,i} C_{k,j})^{-1} (p_{j,i} + P_{j,i} c_{k,j}) + p_{k,j}, \\ 
P_{k,i} &= A_{k,j}^\top (I + P_{j,i} C_{k,j})^{-1} P_{j,i} A_{k,j} + P_{k,j}.
\end{split}  
\end{equation}
Note that $(I + C_{k,j} P_{j,i})^{-1}$ is the transpose of $(I + P_{j,i} C_{k,j})^{-1}$, indicating that only a single LU factorization is required to compute the matrices in \eqref{eq:comb_rule}. 
Moreover, $I + C_{k,j} P_{j,i}$ is nonsingular, provided that both $C_{k,j}$ and $P_{j,i}$ are symmetric positive semidefinite.
Derivation details are provided in Appendices~\ref{app:dual_value_function} and~\ref{app:combination_rule}.

In an LQR problem, we can interpret the one-step CVF $V_{k\to k+1}$ as the input array element $a_k$ in \eqref{eq:array_element}. 
In this setting, the input to the parallel scan algorithm consists of the one-step CVFs followed by the terminal value function, $\left[ V_{0\to 1}, V_{1\to 2}, \dots, V_{N} \right]$, and the corresponding output is the collection of the value functions, $\left[ V_{0}, V_{1}, \dots, V_{N} \right]$.
The associative operator $\oplus$ used to combine two elements is defined by \eqref{eq:V_ki_min}.
Given the input elements and the corresponding operator, we can perform the parallel scan and then determine the affine control law in $O(1)$ parallel time:
\begin{align} \label{eq:feedback_laws}
    K_k &= -\left(R_k + B_k^\top P_{k+1} B_k\right)^{-1} \left( M_k + B_k^\top P_{k+1} A_k \right), \\
    k_k &= -\left(R_k + B_k^\top P_{k+1} B_k\right)^{-1} \left[ r_k + B_k^\top \left( p_{k+1} + P_{k+1}c_k \right) \right]. \nonumber
\end{align}
Next, we conduct a linear rollout to obtain the optimal state and input trajectories using all the feedback gains and affine terms derived in \eqref{eq:feedback_laws}:
\begin{align*}
    x_{k+1}^* &= A_k x_k^* + B_k u_k^* + c_k \\
    &= (A_k + B_k K_k) x_k^* + c_k+ B_k k_k \\
    &= \tilde{A}_k x_k^* + \tilde{c}_k := f_k^{\text{lin}}(x_k^*),
\end{align*} 
where $\tilde{A}_k = A_k + B_k K_k$ and $\tilde{c}_k = c_k+ B_k k_k $.
Due to the affine structure, the parallel scan can also be applied to accelerate this process.
To derive the combination rule, 
we express $x_i^*$ in terms of $x_k^*$ for $i > k$, yielding $x_i^* = f_{k\to i}^{\text{lin}}(x_k^*) = \tilde{A}_{k,i} x_k^* + \tilde{c}_{k,i}$.
The computation of $\tilde{A}_{k,i}$ and $\tilde{c}_{k,i}$ can be split into two parts, forming the combination rule for the forward pass:
\begin{align} \label{eq:forward_comb_rule}
    x_i^* &= f_{k\to j}^{\text{lin}}(x_k^*) \oplus f_{j\to i}^{\text{lin}}(x_j^*) = \tilde{A}_{j,i} ( \tilde{A}_{k,j} x_k^* + \tilde{c}_{k,j}) + \tilde{c}_{j,i},
\end{align}
where $\tilde{A}_{k,i} = \tilde{A}_{j,i}\tilde{A}_{k,j}$ and $\tilde{c}_{k,i} = \tilde{A}_{j,i} \tilde{c}_{k,j} + \tilde{c}_{j,i}$.
Next, we define the input array for the parallel scan using the one-step dynamics:
$
    \left[f_{0\to 1}^{\text{lin}}, f_{1\to 2}^{\text{lin}}, \dots, f_{N-1\to N}^{\text{lin}} \right],
$
where for $k \in [0,N-1]$, the system matrices in $f_{k\to k+1}^{\text{lin}}$ are given by
$
    \tilde{A}_{k, k+1} = \tilde{A}_k, \; \tilde{c}_{k, k+1} = \tilde{c}_k.
$
The output array, resulting from the parallel scan, is then expressed as
\begin{equation}
    [f_{0\to 1}^{\text{lin}}, f_{0\to 2}^{\text{lin}}, \dots, f_{0\to N}^{\text{lin}}], \label{eq:forward_output}
\end{equation}
where each element is obtained as
\begingroup
\begin{align}
    f_{0\to i}^{\text{lin}}(x)
    &= \bigl(f_{0\to 1}^{\text{lin}} \oplus f_{1\to 2}^{\text{lin}} \oplus \dots \oplus f_{i-1\to i}^{\text{lin}}\bigr)(x) \nonumber\\
    &= \tilde{A}_{0,i}x + \tilde{c}_{0,i}.
\end{align}
\endgroup
Finally, the optimal state trajectory $[x^*_1, x^*_2, \dots, x^*_{N}]$ can be recovered via \eqref{eq:forward_output}.

\subsection{Branch Model Predictive Control}

\begin{figure}[!t]\captionsetup[subfigure]{font=scriptsize}
    \centering
    \subfloat[]{\includegraphics[width=1.0\columnwidth]{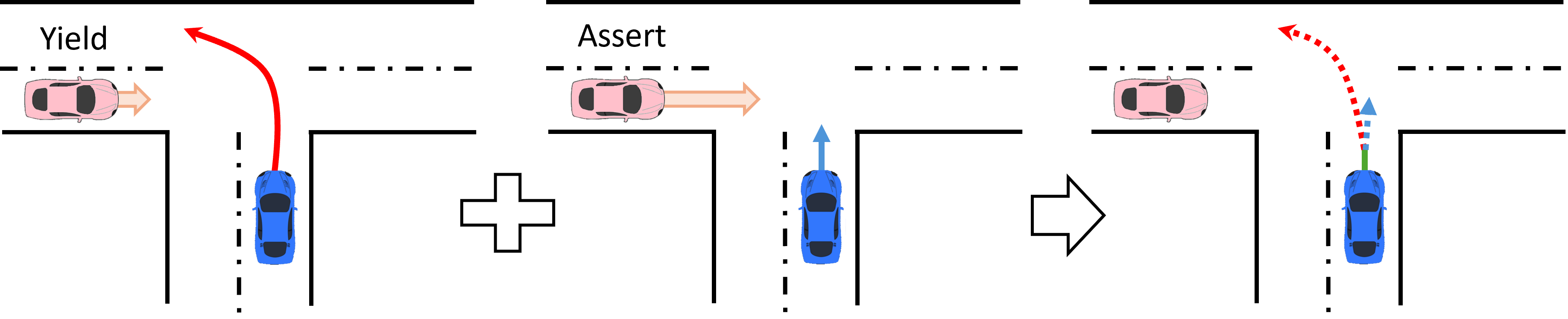}%
    \label{fig:intersection_tree}%
    } \\
    \subfloat[]{\includegraphics[width=0.95\columnwidth]{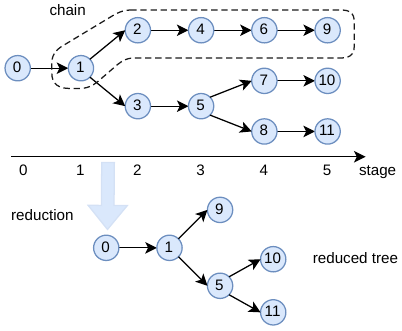}%
    \label{fig:trajectory_tree}%
    }
    \caption{(a) Intersection-crossing scenario.
    The other vehicle (in pink) has two potential driving decisions: ``Yield'' and ``Assert''. 
    The BMPC planner utilizes the trajectory tree to handle both scenarios. 
    (b) Trajectory tree (top) with 12 nodes $\left(\mathcal{N}=\{0, 1, \dots, 11\}\right)$, two branching nodes $\left(\mathcal{B}=\{1, 5\}\right)$, and three leaves $\left(\mathcal{L}=\{9, 10, 11\}\right)$. For node $5$, $\text{p}(5)=3$ and $\text{ch}(5)=\{7, 8\}$.
    A chain is a path whose internal nodes each have a single child; the dashed outline highlights the chain $1\to2\to4\to6\to9$.
    The reduced tree (bottom) replaces each such chain with a single edge while retaining the root, branching nodes, and leaves, giving $\mathcal{N}_{\mathrm{red}}=\{0, 1, 5, 9, 10, 11\}$.
    For example, the highlighted chain becomes the edge $1\to9$ after reduction.}
    \label{fig:tree_example}
\end{figure}

A BMPC planner \cite{zhang_automated_2024} generates a trajectory tree to account for multiple potential scenarios, as shown in Figure \ref{fig:tree_example}. 
In automated driving applications, a scenario can be represented by the driving decision of the surrounding vehicle, as described in Figure \ref{fig:intersection_tree}. 
Specifically, since the ego vehicle (in blue) is uncertain whether the nearby vehicle (in pink) will yield, it plans a trajectory tree with a shared branch that accounts for both possible decisions.
After the branching point, it is assumed that the ego vehicle has complete knowledge of the pink vehicle's decision in this example.
The size of the tree grows rapidly with multiple interacting agents, multi-modal predictions, or multiple branching stages. Large-scale scenario trees also commonly arise in process control problems~\cite{LuciaSMPC2013}.
In Figure~\ref{fig:trajectory_tree}, we represent the index set of all tree nodes as $\mathcal{N}$, where 0 denotes the root node index, the set of all leaf node indices as $\mathcal{L}$, and the set of indices for all nodes at stage $k$ as $\mathcal{N}_k$. 
For tree node $i$, $\text{p}(i)$ denotes the index of its unique direct parent for $i\ne0$, and $\text{ch}(i)$ denotes the index set of its direct children.
Furthermore, we characterize each tree node $i \in \mathcal{N} \backslash \mathcal{L}$ using a tuple $(x_i, u_i, f_i, \ell_i, g_i)$, while a leaf node $i \in \mathcal{L}$ is represented by a tuple $(x_i, \ell_{f,i}, g_i)$.
The state $x_i \in \mathbb{R}^{n_x}$ associated with node $i$ evolves according to the dynamics $x_{j}=f_i(x_i, u_i),\;j\in\text{ch}(i)$, where $f_i: \mathbb{R}^{n_x} \times  \mathbb{R}^{n_u} \rightarrow  \mathbb{R}^{n_x}$ is a general nonlinear function.
Additionally, $\ell_i: \mathbb{R}^{n_x} \times \mathbb{R}^{n_u} \rightarrow \mathbb{R}$ and $\ell_{f,i}: \mathbb{R}^{n_x} \rightarrow \mathbb{R}$ represent the stage and final costs, respectively.
We define the cost function of the BMPC as the sum of the individual node costs and formulate the optimal control problem as follows:
\begin{subequations} \label{eq:bmpc}
\begin{align} 
    \min_{\boldsymbol{x}, \boldsymbol{u}} &\sum_{i \in \mathcal{L}} \ell_{f, i}\left(x_i\right) + 
    \sum_{i \in \mathcal{N} \backslash \mathcal{L}} \ell_i\left(x_i, u_i \right) \label{eq:tree_cost}\\
    \text{s.t.}\; 
    & x_{i} = f_{\text{p}(i)}(x_{\text{p}(i)}, u_{\text{p}(i)}),\;  \forall i \in \mathcal{N} \backslash \{0\}, \quad x_0 = \bar{x}, \label{eq:tree_ev_dyn}\\
    & g_i(x_i, u_i) \leq 0,\; \forall i \in \mathcal{N} \backslash \mathcal{L},\;\; g_i(x_i) \leq 0,\; \forall i \in \mathcal{L}  \label{eq:tree_ineq},
\end{align}
\end{subequations}
where $\boldsymbol{x} := \left( x_i \right)_{i \in \mathcal{N}\backslash\{0\}}$ and $\boldsymbol{u} := \left( u_i \right)_{i \in \mathcal{N} \backslash \mathcal{L}}$ are the state and input trees, $\bar{x}$ is the current measured state, 
and the general inequality constraints are represented by \eqref{eq:tree_ineq}.
We denote the stacked inequality constraints by $\boldsymbol{g}(\boldsymbol{x}, \boldsymbol{u})$.
Node weights, when available, can be incorporated into the node costs.

\section{Parallel Tree LQR} \label{sec:parallel_tree_lqr}
\subsection{Scenario-Level Parallelization}
We begin by introducing a scenario-level parallel method $\textsc{sLQR}$, which requires fewer GPU resources (e.g., CUDA cores).
Let us consider the following tree LQR problem:
\begin{align} \label{eq:lqr_tree_approx}
    \min_{\boldsymbol{x}, \boldsymbol{u}}\; &\sum_{i \in \mathcal{L}} \ell^{\text{qp}}_{f, i} (x_i) + \sum_{i \in \mathcal{N} \backslash \mathcal{L}} \ell^{\text{qp}}_{i} (x_i, u_i)
    \\
    \text{s.t.}\;
    & x_{i} = A_{\text{p}(i)} x_{\text{p}(i)} + B_{\text{p}(i)} u_{\text{p}(i)} + c_i,\;  \forall i \in \mathcal{N} \backslash \{0\} \nonumber,
\end{align} 
where $\ell^{\text{qp}}_{i}$ is the quadratic stage cost as in~\eqref{eq:LQR}, and $c_i$ often represents dynamics defects.
To address~\eqref{eq:lqr_tree_approx}, we perform a modified Riccati recursion from the leaf nodes to the root, followed by a forward pass to recover the optimal trajectory tree. 
Specifically, at stage $k$, we solve $\vert \mathcal{N}_k\vert$ minimization problems: 
\begin{equation} \label{eq:tree_Riccati}
    \begin{aligned}
        \min_{u_n}\; \ell^{\text{qp}}_n(x_n, u_n) + \sum_{m\in \text{ch}(n)} V_{m} (x_m),\; n \in \mathcal{N}_k, 
    \end{aligned}
\end{equation}
where $\vert \mathcal{N}_k\vert$ represents the number of tree nodes at stage $k$, and $V_{m}(\cdot)$ denotes the value function (VF) at node $m$.
Unlike the classic Riccati recursion, the modified recursion accounts for the tree structure by aggregating the value functions of all child nodes, rather than using a single value function.
Once the child value functions are available, the $\vert\mathcal{N}_k\vert$ optimization problems in~\eqref{eq:tree_Riccati} are independent and can be solved in parallel. For example, in Figure~\ref{fig:trajectory_tree}, the problems corresponding to nodes 4 and 5 at stage 3 can be solved in parallel.

\subsection{Scenario and Temporal Parallelization}
When GPU resources are sufficient, we employ the parallel scan to enable scenario-level and temporal parallelism in both the backward and forward passes $(\textsc{stLQR})$, improving computational efficiency.
We begin by introducing the reduced tree, as shown in Figure~\ref{fig:trajectory_tree}.
Let $\mathcal B$ and $\mathcal N_{\mathrm{red}}$ denote the set of branching nodes and the set of nodes in the reduced tree:
\begin{equation}\label{eq:reduced_nodes}
\begin{aligned}
\mathcal B = \left\{ n \in \mathcal N \mid |\text{ch}(n)|\ge2 \right\},\;\;
\mathcal N_{\mathrm{red}} = \{0\}\cup\mathcal B\cup\mathcal L.
\end{aligned}
\end{equation}
For a reduced edge from $n\in\mathcal N_{\mathrm{red}}\setminus\mathcal L$ to $m\in\mathcal N_{\mathrm{red}}$, the original tree path is represented as $n\to s\to\cdots\to m$, where $s\in\text{ch}(n)$.
For each reduced edge, we retain the stage cost and control input $u_n$ at node $n$, together with the first transition $x_s=A_nx_n + B_nu_n + c_s$.
We use the chain CVF $V_{s\to m}(x_s,x_m)$ to account for the internal stages from $s$ to $\text{p}(m)$. The stage cost at $m$ is excluded from this CVF.
For example, the highlighted path $1\to2\to4\to6\to9$ in Figure~\ref{fig:trajectory_tree} corresponds to the reduced edge $1\to9$, with $n=1$, $s=2$, and $m=9$.
The stage cost and control at node $1$, together with the transition $1\to2$, remain explicit, while $V_{2\to9}(x_2,x_9)$ accounts for the stages at nodes $2$, $4$, and $6$.

For each chain to be reduced, we construct its CVFs $[V_{s\to m}, \dots, V_{\text{p}(m)\to m}]$ using a parallel scan based on~\eqref{eq:comb_rule}. 
These computations are independent across paths and can therefore be performed in parallel. 
Once the VF $V_m$ at node $m$ is available, the VF at node $s$ is obtained as 
\begin{equation} \label{eq:V_s}
    V_s(x_s)=\min_{x_m}\{V_{s\to m}(x_s,x_m)+V_m(x_m)\}.
\end{equation}
Using the first-order optimality conditions of the minimization problem~\eqref{eq:V_s}, we express the optimal state at node $m$ as $x_m^*=\tilde{A}_{s,m}x_s+\tilde{c}_{s,m}$, with the coefficients determined by
\begin{equation}\label{eq:chain_condition}
\begin{aligned}
    E_{s,m}&=I+C_{s,m}P_m,\\
    E_{s,m}\tilde{A}_{s,m}&=A_{s,m},\\
    E_{s,m}\tilde{c}_{s,m}&=c_{s,m}-C_{s,m}p_m.
\end{aligned}
\end{equation}
Differentiating the optimal value in~\eqref{eq:V_s} with respect to $x_s$ and matching coefficients gives the quadratic and linear coefficients of $V_s$:
\begin{equation}\label{eq:chain_inlet_value}
\begin{aligned}
P_s&=P_{s,m}+A_{s,m}^\top P_m\tilde{A}_{s,m},\\
p_s&=p_{s,m}+A_{s,m}^\top(p_m+P_m\tilde{c}_{s,m}).
\end{aligned}
\end{equation}

We provide the detailed derivations of~\eqref{eq:chain_condition} and~\eqref{eq:chain_inlet_value} in Appendix~\ref{app:chain_value_recovery}.
Once the VF $V_s$ has been computed for every child $s\in\text{ch}(n)$, we calculate the VF associated with the retained node $n$ based on~\eqref{eq:tree_Riccati}.
All child contributions are summed before eliminating $u_n$.
Retained nodes with the same subtree height in the reduced tree can be processed in parallel. 
These nodes may stay at different stages of the original tree.

At the end of the reduction phase, the value functions $V_m$ are available for all $m \in \mathcal{N}_{\mathrm{red}}$.
We next recover the VFs at the internal chain nodes from the previously computed CVFs using~\eqref{eq:chain_condition}--\eqref{eq:chain_inlet_value}.
For a chain edge $i\to j$, $K_i,k_i$ are obtained from~\eqref{eq:feedback_laws} with offset $c_j$.

In the forward pass, starting from the given $x_0^*$, we propagate the state along each reduced edge $n\to s\to\cdots\to m$ as follows:
\begin{equation}\label{eq:reduced_forward}
\begin{aligned}
u_n^*&=K_nx_n^*+k_n,\\
x_s^*&=A_nx_n^*+B_nu_n^* + c_s,\\
x_m^*&=\tilde{A}_{s,m}x_s^*+\tilde{c}_{s,m}.
\end{aligned}
\end{equation}
We process retained nodes at the same depth in the reduced tree in parallel. 
Once the initial internal state $x_s^*$ of each chain is available, based on the affine composition rule~\eqref{eq:forward_comb_rule}, independent affine scans recover the internal states and inputs: $x_i^*=\tilde{A}_{s,i}x_s^*+\tilde{c}_{s,i}$ and
$u_i^*=K_ix_i^*+k_i$.

\section{Parallel Branch MPC} \label{sec:parallel_bmpc}
This section presents an efficient parallel solver for BMPC problems. We adopt the augmented Lagrangian (AL) method to handle stage-wise constraints \eqref{eq:tree_ineq}. 
Compared to interior-point methods, AL-based methods can be easily warm-started, making them well suited for MPC problems~\cite{jallet_parallel_2024}.  
While SQP methods can also benefit from warm-starting, their algorithmic framework is often more complex than that of AL-based methods due to the feasibility issues introduced by linearization~\cite{GillSNOPT2002}.
Next, we describe the nested loops in our AL-based method.

\subsection{Inner Loop}
In the inner loop, stage-wise constraints are penalized using the Powell-Hestenes-Rockafellar (PHR) function~\cite{rockafellar1976augmented}:
\begingroup\small
\begin{equation} \label{eq:AL}
\begin{aligned}
    &L_a(\boldsymbol{x},\boldsymbol{u};\boldsymbol{y}_e,\rho)\\
    &=\sum_{i\in\mathcal{L}}\Bigl[\ell_{f,i}(x_i)
      +\tfrac{\rho}{2}\bigl\|\max\bigl(0,g_i(x_i)+\tfrac{y_{e,i}}{\rho}\bigr)\bigr\|^2\Bigr]\\
    &\mathrel{\phantom{=}}+\sum_{i\in\mathcal{N}\backslash\mathcal{L}}\Bigl[\ell_i(x_i,u_i)
      +\tfrac{\rho}{2}\bigl\|\max\bigl(0,g_i(x_i,u_i)+\tfrac{y_{e,i}}{\rho}\bigr)\bigr\|^2\Bigr].
\end{aligned}
\end{equation}
\endgroup
where $y_{e,i}$ denotes the estimated Lagrange multiplier, and $\rho > 0$ is the penalty weight.
The relaxed optimal control problem for \eqref{eq:bmpc} is then formulated as
\begingroup
\begin{equation} \label{eq:relax_bmpc}
\begin{aligned} 
    \min_{\boldsymbol{x}, \boldsymbol{u}}\quad & L_a(\boldsymbol{x}, \boldsymbol{u}; \boldsymbol{y}_e, \rho) \\
    \text{s.t.}\quad
    & x_0 = \bar{x}, \\
    & x_{i} = f_{\text{p}(i)}\left(x_{\text{p}(i)}, u_{\text{p}(i)}\right),\;  \forall i \in \mathcal{N} \backslash \{0\},
\end{aligned}
\end{equation}
\endgroup
where the penalty weight and the Lagrange multipliers are held fixed in the inner loop and updated in the outer loop.
\subsubsection{Search Direction} 
To address \eqref{eq:relax_bmpc}, we adopt an iLQR-type method. 
At each iteration, a tree LQR problem is constructed by approximating \eqref{eq:relax_bmpc} around the trajectory tree obtained from the previous iteration.
The cost matrices in \eqref{eq:lqr_tree_approx} are given by the gradients and Hessian matrices of the Lagrangian corresponding to \eqref{eq:relax_bmpc}:
\begin{align*}
    \mathcal{L}(\boldsymbol{x}, \boldsymbol{u}, \boldsymbol{\lambda}) &=  L_a(\boldsymbol{x}, \boldsymbol{u}; \boldsymbol{y}_e, \rho) \\
    &\qquad + \sum_{i \in \mathcal{N} \backslash \{0\}}   \lambda_i^\top \left(f_{\text{p}(i)}(x_{\text{p}(i)}, u_{\text{p}(i)}) - x_{i} \right), 
\end{align*}
with $\lambda_i \in \mathbb{R}^{n_x}$ representing the Lagrange multiplier for the dynamic constraint.
To ensure that the approximated Hessian for \eqref{eq:relax_bmpc} is positive semidefinite, we use either the generalized Gauss-Newton Hessian approximation~\cite{schraudolph_fast_2002, HowellALTRO2019, GrandiaLocomotion2023} or the exact Hessian with eigenvalue modification~\cite{Nocedal_numerical_2006}. 
At each inner iteration, the resulting tree LQR problem is solved for the state and input perturbations using the parallel method introduced in Section~\ref{sec:parallel_tree_lqr}.

In an iLQR-type solver, a nonlinear rollout is commonly used \cite{HowellALTRO2019, TassaiLQR2012}. 
However, without an affine structure, nonlinear rollouts are difficult to parallelize and therefore run inefficiently on GPUs. 
Thus, we employ the parallel linear rollout to compute the optimal perturbations of the state and input trees, $(\delta \hat{\boldsymbol{x}}, \delta \hat{\boldsymbol{u}})$.
Our method is classified as a multiple-shooting (MS) approach due to the use of the linear rollout.

\subsubsection{Line Search}
After the rollout, we execute a line search to determine the best step size $\alpha \in (0, 1]$. 
As the linear rollout cannot ensure dynamic feasibility, we adopt the augmented Lagrangian merit function to balance cost reduction and constraint violations \cite{GillSNOPT2002}:
\begin{equation} \label{eq:merit}
\begin{aligned}
    &M(\boldsymbol{x}, \boldsymbol{u}; \mu) = L_a(\boldsymbol{x}, \boldsymbol{u}; \boldsymbol{y}_e, \rho) \\
    &\quad + \boldsymbol{\lambda}^\top \boldsymbol{d}(\boldsymbol{x}, \boldsymbol{u}) + \dfrac{\mu}{2} \boldsymbol{d}(\boldsymbol{x}, \boldsymbol{u})^\top \boldsymbol{d}(\boldsymbol{x}, \boldsymbol{u}),
\end{aligned}
\end{equation}
where $\mu > 0$ is a penalty parameter, and $\boldsymbol{d}(\cdot, \cdot)$ collects the dynamics residuals. 
Compared with the nonsmooth $\ell_1$ merit function used in \cite{LiUnified2023}, the augmented
Lagrangian avoids the Maratos effect, resulting in faster local convergence \cite{GillSQP2012}.
We decide if the trial step $(\alpha \delta \hat{\boldsymbol{x}}, \alpha \delta \hat{\boldsymbol{u}})$ can be accepted based on the following sufficient decrease condition:
\begin{equation} \label{eq:sufficient_decrease}
    \phi(\alpha;\mu) \leq \phi(0;\mu) + \beta \alpha \phi^\prime(0;\mu),
\end{equation}
where $\beta \in (0, 1)$, $\phi(\alpha;\mu) := M(\hat{\boldsymbol{x}} + \alpha\delta \hat{\boldsymbol{x}}, \hat{\boldsymbol{u}} + \alpha\delta \hat{\boldsymbol{u}}; \mu)$ is the line search function, and $\phi^\prime(\alpha;\mu)$ denotes its derivative.
Following \cite{GillAugLag1992}, we adopt an adaptive update method for $\mu$ to ensure that the reduction $\phi^\prime(0;\mu)$ is sufficient. 
The computation of trial steps is easily parallelizable.
Once all acceptable step sizes are obtained according to \eqref{eq:sufficient_decrease}, we select the largest one following the backtracking approach in \cite{Nocedal_numerical_2006}.

\subsubsection{Inner Loop Stopping Criterion}
We use the following KKT residual associated with \eqref{eq:relax_bmpc} as the convergence metric:
\begin{equation*}
    r_{\text{kkt}} := \max\left\{ \left\Vert\nabla_{\boldsymbol{x},\boldsymbol{u}}\mathcal{L} \right\Vert_{\infty}, \left\Vert\nabla_{\boldsymbol{\lambda}}\mathcal{L} \right\Vert_{\infty} \right\}.
\end{equation*}
In the beginning, solving the inner minimization exactly is unnecessary since $\boldsymbol{y}_e$ may not yet provide a reliable estimate. 
Therefore, the inner loop terminates when $r_{\text{kkt}} \leq \omega$, for a prescribed tolerance $\omega$. A heuristic for choosing $\omega$ is provided in the following section.

\subsection{Outer Loop}
In the outer loop, we adopt the BCL rule proposed in \cite{ConnBCL1991} and implemented in \cite{JalletProxDDP2025}. This heuristic adaptively updates the dual variables $\boldsymbol{y}_e$ and the penalty weight $\rho$ based on the constraint violation $c := \left\Vert \max\left( \boldsymbol{g}(\boldsymbol{x}, \boldsymbol{u}), -\boldsymbol{y}_e / \rho \right) \right\Vert_{\infty}$.
Specifically, once the inner loop terminates, we evaluate the constraint violation $c^{k+1}$. If $c^{k+1}$ fails to decrease below $\eta^k$, we increase the penalty weight to $\rho^{k+1} = \tau \rho^k$, encouraging the next iteration to place greater emphasis on constraint satisfaction. 
If $c^{k+1} \leq \eta^k$, we update the dual variables and keep the penalty weight unchanged. Additional technical details can be found in Algorithm \ref{alg:alg1}, including the scheduling of tolerances. The $\max$ operator is applied componentwise.
The algorithm is deemed converged when $\max\left( r^k_{\text{kkt}}, c^{k+1} \right) < \epsilon_{\text{tol}}$. 

\begin{algorithm}[H]
\caption{AL Method}\label{alg:alg1}
\begin{algorithmic}[1]
\While{Maximum number of iterations not reached}
    \State $\hat{\boldsymbol{x}}, \hat{\boldsymbol{u}} \leftarrow \boldsymbol{x}^k, \boldsymbol{u}^k$
    \While{$r^k_{\text{kkt}} > \omega^k$}
        \State Linearize the dynamics and quadratize
        \Statex \hspace{\algorithmicindent\quad} the objective around $(\hat{\boldsymbol{x}}, \hat{\boldsymbol{u}})$ to construct~\eqref{eq:lqr_tree_approx}
        \State $\delta \hat{\boldsymbol{x}}, \delta \hat{\boldsymbol{u}} \leftarrow $ solve Tree LQR  using~\eqref{eq:chain_condition},~\eqref{eq:chain_inlet_value},~\eqref{eq:reduced_forward}
        \State Compute $\mu^k$ according to \cite{GillAugLag1992}
        \State Compute $\phi(0; \mu^k)$ and $\phi^\prime(0; \mu^k)$
        \ParFor{all $\alpha_i$}
            \State $(\hat{\boldsymbol{x}}^\prime, \hat{\boldsymbol{u}}^\prime) \leftarrow (\hat{\boldsymbol{x}}, \hat{\boldsymbol{u}}) + \alpha_i (\delta \hat{\boldsymbol{x}}, \delta \hat{\boldsymbol{u}})$
            \State Evaluate the line search function $\phi(\alpha_i; \mu^k)$
        \EndParFor 
        \State Find $\{\alpha_i^\prime\} \subseteq \{\alpha_i\}$ that satisfies 
        \eqref{eq:sufficient_decrease}
        \State $\alpha^* \gets \max\; \{\alpha_i^\prime\}$
        \State $(\hat{\boldsymbol{x}}, \hat{\boldsymbol{u}}) \leftarrow (\hat{\boldsymbol{x}}, \hat{\boldsymbol{u}}) + \alpha^* (\delta \hat{\boldsymbol{x}}, \delta \hat{\boldsymbol{u}})$
    \EndWhile
    \State $\boldsymbol{x}^{k+1}, \boldsymbol{u}^{k+1} \leftarrow \hat{\boldsymbol{x}}, \hat{\boldsymbol{u}}$ 
    \State $c^{k+1} \leftarrow \left\Vert \max\left( \boldsymbol{g}(\boldsymbol{x}^{k+1}, \boldsymbol{u}^{k+1}), -\boldsymbol{y}_e^k / \rho^k \right) \right\Vert_{\infty}$ 
    \If{$c^{k+1} \leq \eta^k$}
        \State $\boldsymbol{y}^{k+1}_e \leftarrow \max\left( 0, \boldsymbol{y}_e^k + {} \rho^k \boldsymbol{g}(\boldsymbol{x}^{k+1}, \boldsymbol{u}^{k+1}) \right)$,
        \Statex \hspace{\dimexpr\algorithmicindent+\algorithmicindent\relax} $\rho^{k+1} \leftarrow \rho^k$
        \If{$\max\left( r^k_{\text{kkt}}, c^{k+1} \right) < \epsilon_{\text{tol}}$}
             \State \Return $\boldsymbol{x}^{k+1}, \boldsymbol{u}^{k+1}, \boldsymbol{y}^{k+1}_e$
        \EndIf
        \State $\omega^{k+1} \leftarrow \omega^{k} (\rho^{k+1})^{-\beta_{\omega}}$ ,\, $\eta^{k+1} \leftarrow \eta^{k} (\rho^{k+1})^{-\beta_{\eta}}$
    \Else{}
        \State $\boldsymbol{y}_e^{k+1} \leftarrow \boldsymbol{y}_e^{k}$,\, $\rho^{k+1} \leftarrow \tau \rho^k$
        \State $\omega^{k+1} \leftarrow \omega^{0} (\rho^{k+1})^{-\alpha_{\omega}}$ ,\, $\eta^{k+1} \leftarrow \eta^{0} (\rho^{k+1})^{-\alpha_{\eta}}$
    \EndIf
    \State $\omega^{k+1} \leftarrow \max\left(\omega^{k+1}, \epsilon_{\text{tol}} \right),\, k \leftarrow k + 1$  
\EndWhile
\end{algorithmic}
\label{alg1}
\end{algorithm}

\section{Numerical Experiments} \label{sec:results}

We implement the proposed solver using \textsc{JAX}~\cite{jax2018github}.
All simulations are conducted on a desktop with an 
Intel Core i5-14600KF processor and an NVIDIA GeForce RTX 5060 Ti.
The GPU supports both single-precision (FP32) and double-precision (FP64) arithmetic. Unless stated otherwise, we use FP64 in the numerical experiments for better numerical stability.
The code will be publicly available at \url{https://github.com/Luyao787/ilqrx}.

\subsection{Tree LQR} \label{subsec:exp_tree_lqr}
\begin{figure}[!t]\captionsetup[subfigure]{font=scriptsize}
    \centering
    \subfloat[]{\includegraphics[width=0.35\columnwidth]{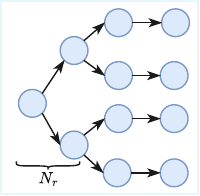}%
    \label{fig:tree_exp}
    }
    \hspace{7.5mm}
    \subfloat[]{\includegraphics[width=0.35\columnwidth]{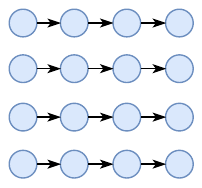}%
    \label{fig:batch_exp}    
    }
    \caption{(a) Tree example used in numerical experiments with $N_r=2$ and $n_d=2$. (b) Batch of trajectories.}
\end{figure}

In this section, we evaluate the computational performance of the proposed tree LQR solvers: $\textsc{sLQR}$ and $\textsc{stLQR}$.  
Additionally, we compare our approaches against the tree LQR solver in $\textsc{HPIPM}$~\cite{FRISON2020HPIPM}.
To conveniently vary the problem size, including the horizon length, state dimension, and number of tree paths, we consider a chain of masses and springs, as in \cite{FRISON201714399}. 
The system consists of $M$ masses, where each mass is characterized by its position and velocity, resulting in $n_x = 2 M$ and $n_u = M-1$.
We assume that model mismatch is due to uncertainty in the spring constant.
An example of the tree structure is illustrated in Figure~\ref{fig:tree_exp}, where the robust horizon is denoted by $N_r$.
We sample $n_d$ spring constants, and the corresponding number of tree paths is $Z = n_d^{N_r}$. 

\subsubsection{Single LQR}
\begin{figure}[!htbp]
    \centering
    \includegraphics[width=0.9\linewidth]{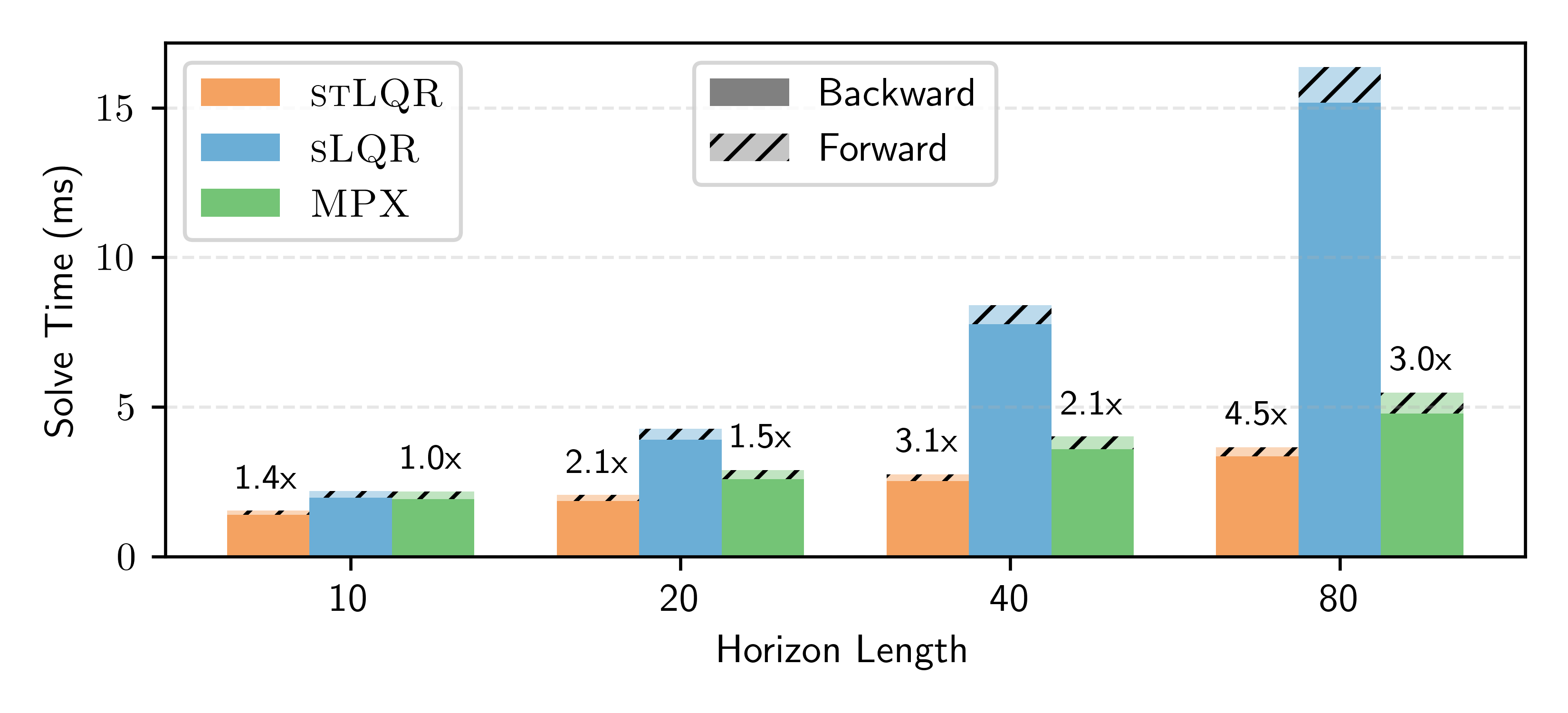}
    \caption{Solve time versus horizon length. Labels above the bars show the speedup relative to $\textsc{sLQR}$.}
    \label{fig:lqr_horizon}
\end{figure}

We set $N_r=1$ and $n_d=1$, and fix the state dimension at $n_x=30$, resulting in an LQR problem without a tree structure.
As shown in \cite{SarkkaTemporal2023}, the parallel-scan–based LQR solver outperforms the Riccati-based solver on GPUs for long-horizon problems.
We additionally assess its performance on medium-horizon problems (e.g., $N \leq 100$) and report the corresponding results in Figure~\ref{fig:lqr_horizon}. 
We observe that the parallel solver remains faster than its serial counterpart, even when the horizon length is short ($N \leq 20$), possibly because the serial method cannot effectively hide GPU memory-access latency.
Moreover, we compare against the parallel LQR solver in $\textsc{MPX}$ \cite{AmatucciMPX2026}, finding that our $\textsc{stLQR}$ is more efficient.
Since our focus is on GPU-oriented implementations, \textsc{HPIPM} is not included in Figure~\ref{fig:lqr_horizon}. For medium-horizon LQR problems, \textsc{HPIPM} can outperform the GPU-based methods considered here due to GPU overhead and the current limitations of the \textsc{JAX}-based implementation.

\subsubsection{Batch LQR}
\begin{figure*}[!t]\captionsetup[subfigure]{font=scriptsize}
    \centering
    \subfloat[]{\includegraphics[width=0.6\columnwidth]{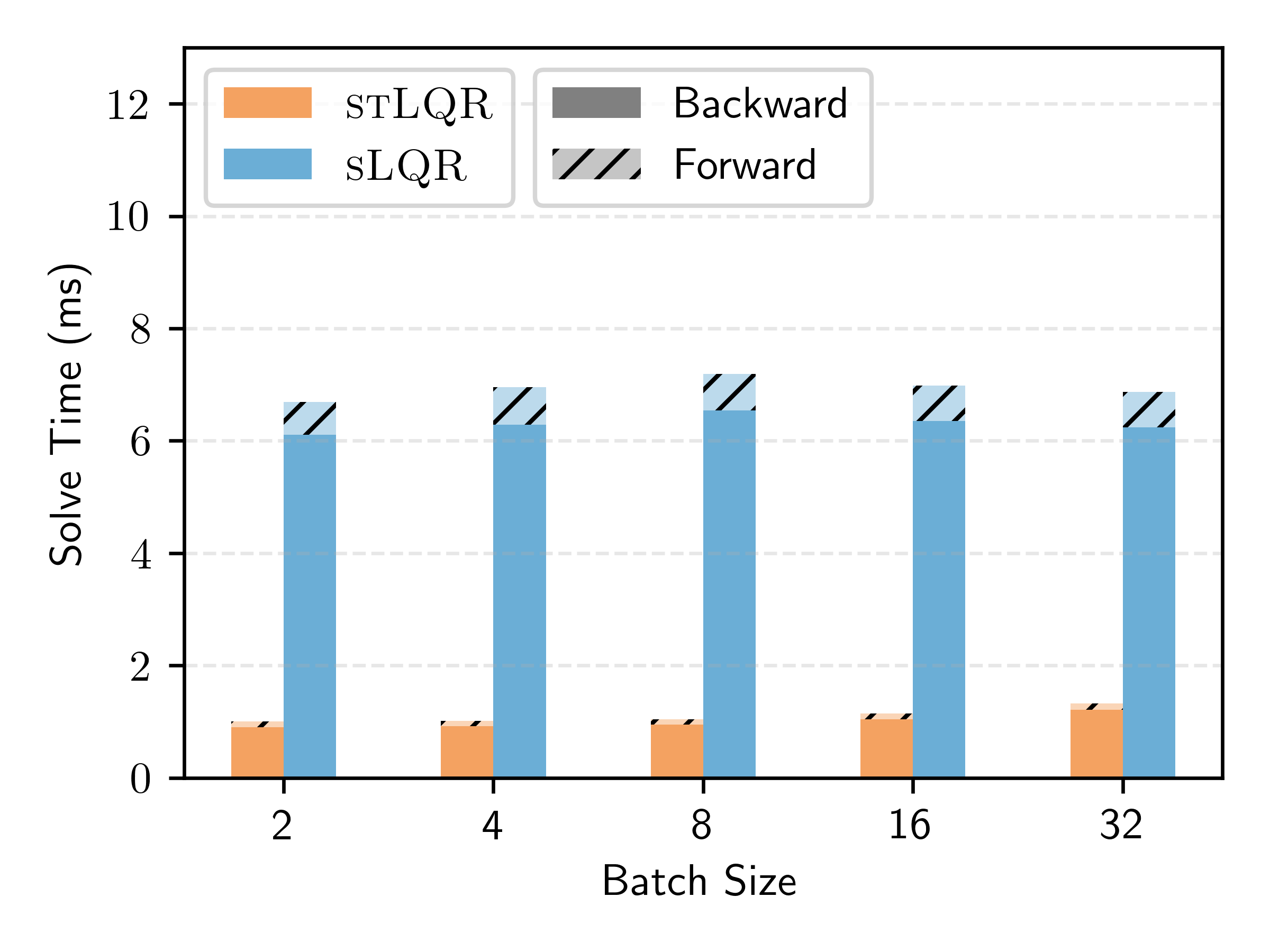}%
    \label{fig:lqr_batch_5_FP32}
    }
    \subfloat[]{\includegraphics[width=0.6\columnwidth]{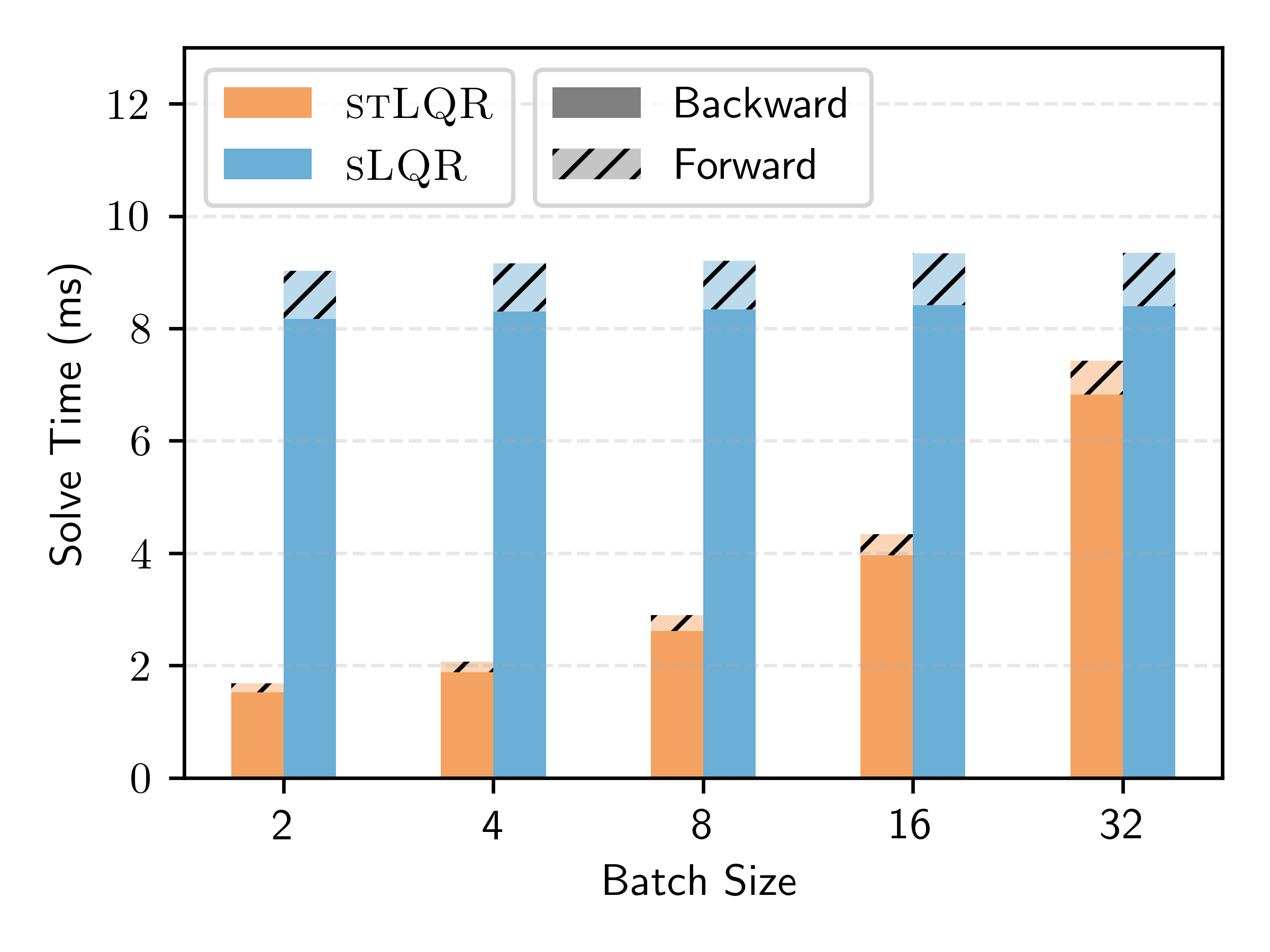}%
    \label{fig:lqr_batch_5_FP64}    
    }
    \subfloat[]{\includegraphics[width=0.6\columnwidth]{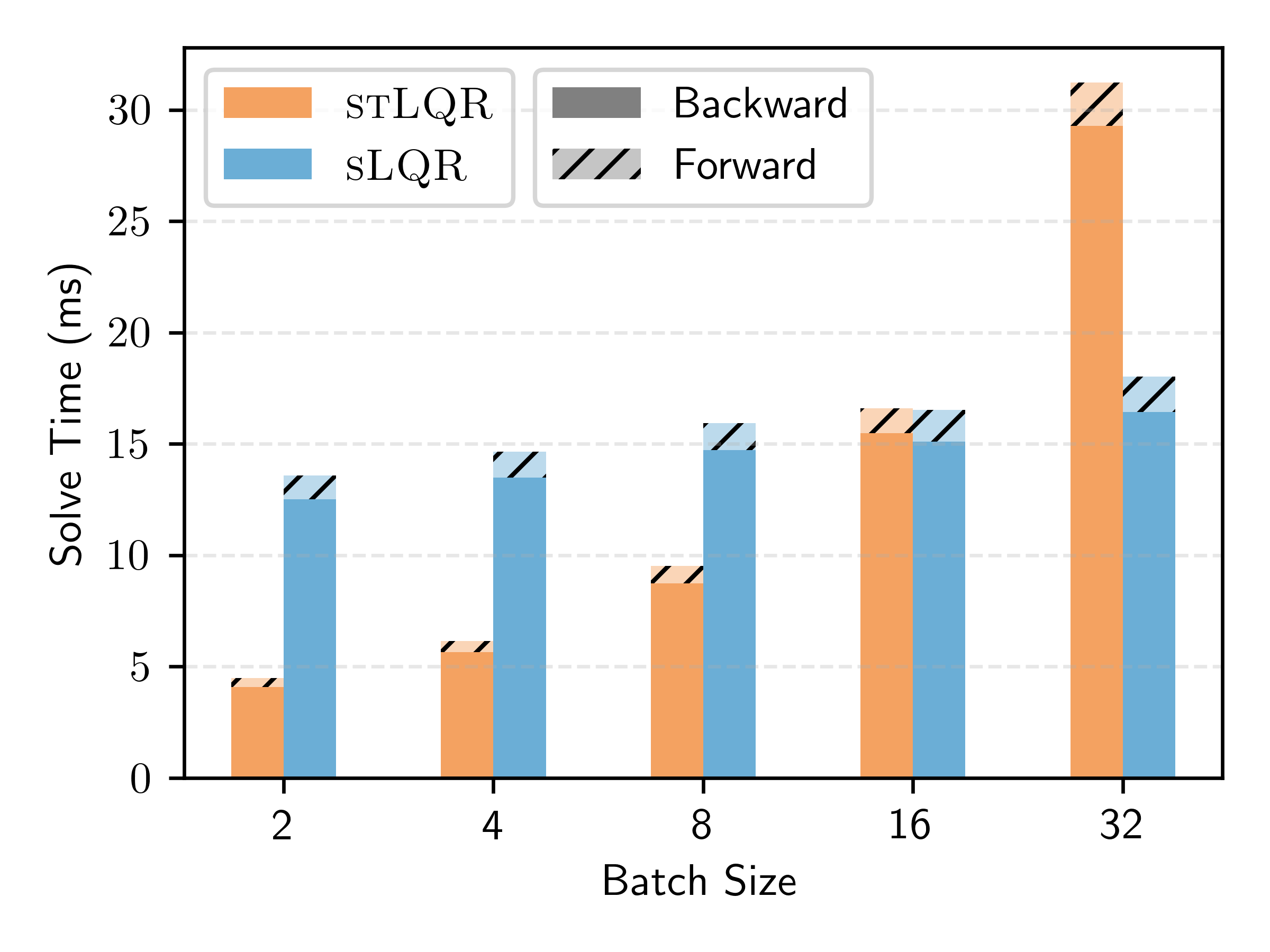}%
    \label{fig:lqr_batch_15_FP64}    
    }
    \caption{Solve time versus batch size. (a) State dimension $n_x=10$ with FP32. (b) State dimension $n_x=10$ with FP64. (c) State dimension $n_x=30$ with FP64.}
    \label{fig:lqr_batch_5}
\end{figure*}

Next, we investigate batch LQR computation, which is closely related to tree LQR problems.
We begin with a relatively small state dimension $(n_x=10)$ and use FP32 to save GPU resources. The horizon length is fixed at $N=80$.
Solve times for different solvers at varying batch sizes are reported in Figure~\ref{fig:lqr_batch_5_FP32}. Both solvers scale well with batch size since the GPU resources are not yet saturated.
We then use FP64 and show the results in Figure~\ref{fig:lqr_batch_5_FP64}. 
Relative to FP32, the FP64 solvers show longer solve times, as the GPU used in the experiment has low FP64 throughput. 
Moreover, the solve time of \textsc{stLQR} increases noticeably with batch size, indicating saturation of GPU resources.
To further increase the computational load, we set the state dimension to $n_x=30$.
In Figure~\ref{fig:lqr_batch_15_FP64}, we observe that the solve time of \textsc{sLQR} increases only slightly with batch size, whereas \textsc{stLQR} scales less favorably: when the batch size increases from 16 to 32, its solve time roughly doubles.

\subsubsection{Tree LQR}
\begin{figure}[!htbp]
    \centering
    \includegraphics[width=0.85\linewidth]{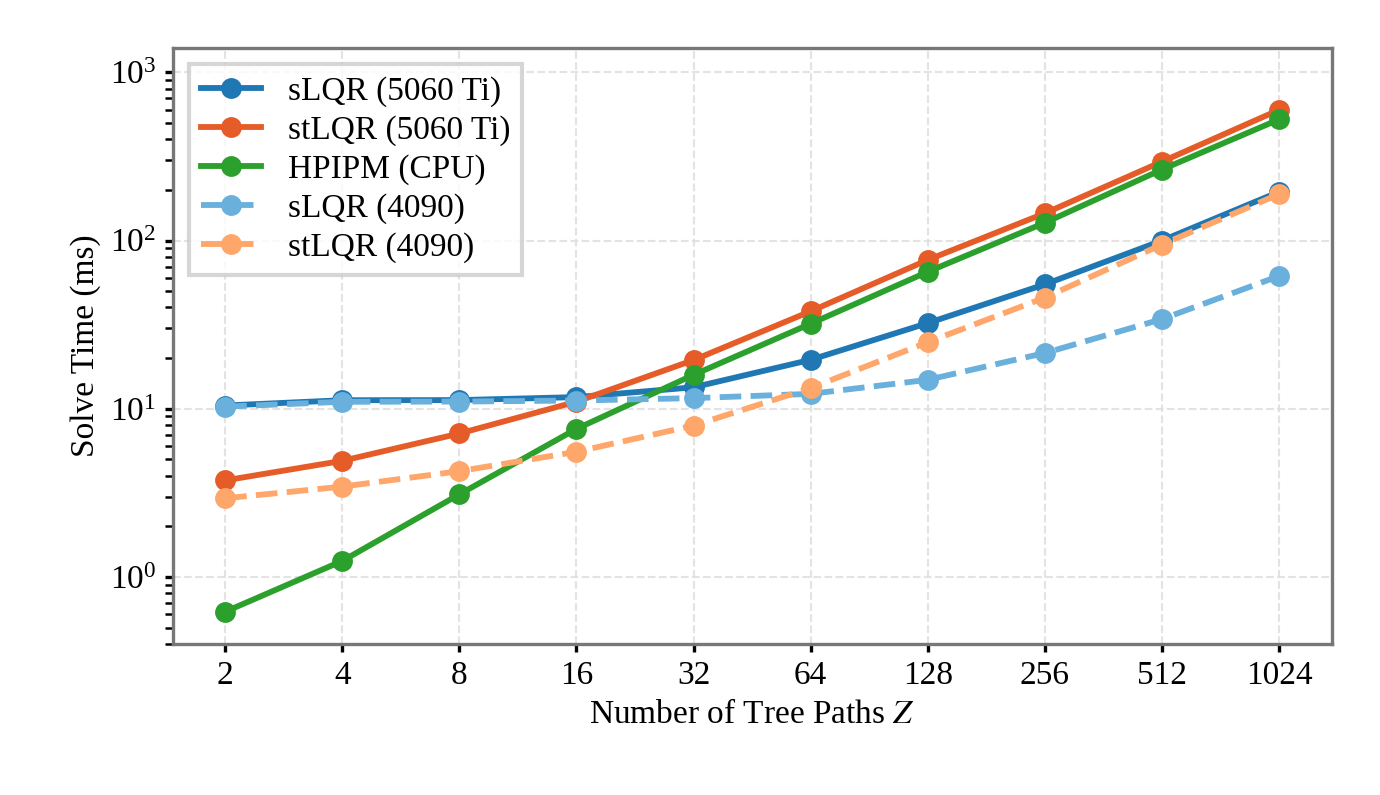}
    \caption{Solve time versus number of tree paths. The corresponding robust horizon length varies from  1 to 10.
    The labels `4090' and `5060 Ti' refer to the GPU models (NVIDIA GeForce RTX 4090/5060 Ti).}
    \label{fig:tree_lqr_treepath}
\end{figure}

We set $n_d=2$ and vary the robust horizon length $N_r$ to obtain different numbers of tree paths. The state dimension is fixed at $n_x=30$. 
As illustrated in Figure~\ref{fig:tree_lqr_treepath}, when $Z=2$, the GPU-based solvers $\textsc{stLQR}$ and $\textsc{sLQR}$ are over $6\times$ and $17\times$ slower than the CPU-based $\textsc{HPIPM}$, respectively. 
A major reason is GPU memory-access latency, along with the additional overhead of data transfers and kernel launches.
For $\textsc{stLQR}$, the solve time grows approximately linearly with the number of tree paths for $Z\geq8$, indicating GPU resource saturation.
In contrast, $\textsc{sLQR}$ saturates at a larger problem size and can outperform $\textsc{HPIPM}$ by up to $2.7\times$ on large-scale instances $(Z \geq 128)$.
We further benchmark our solvers on a high-end GPU: the NVIDIA GeForce RTX 4090. 
We observe that the performance of both solvers begins to saturate at a later point. 
With sufficient GPU resources, $\textsc{stLQR}$ can outperform $\textsc{HPIPM}$ when the number of tree paths is moderate or larger $(Z \geq 16)$, achieving up to a $2.8\times$ speedup, while $\textsc{sLQR}$ remains the best choice when the tree path count is large $(Z \geq 128)$.



\subsection{Constraint Handling}
\begin{figure}[!htbp]\captionsetup[subfigure]{font=scriptsize}
    \centering
    \subfloat[]{\includegraphics[width=0.49\columnwidth]{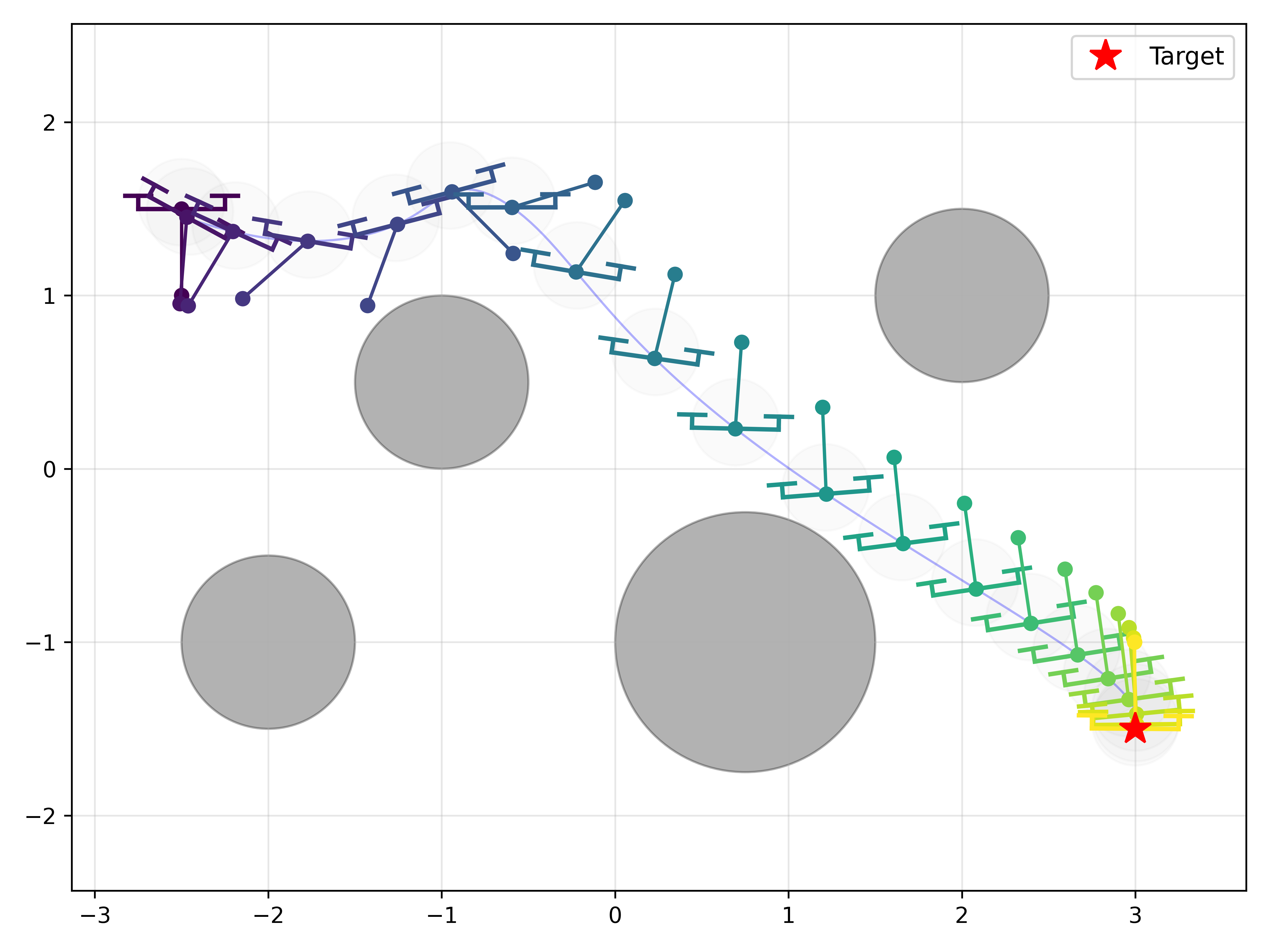}%
    }
    \subfloat[]{\includegraphics[width=0.49\columnwidth]{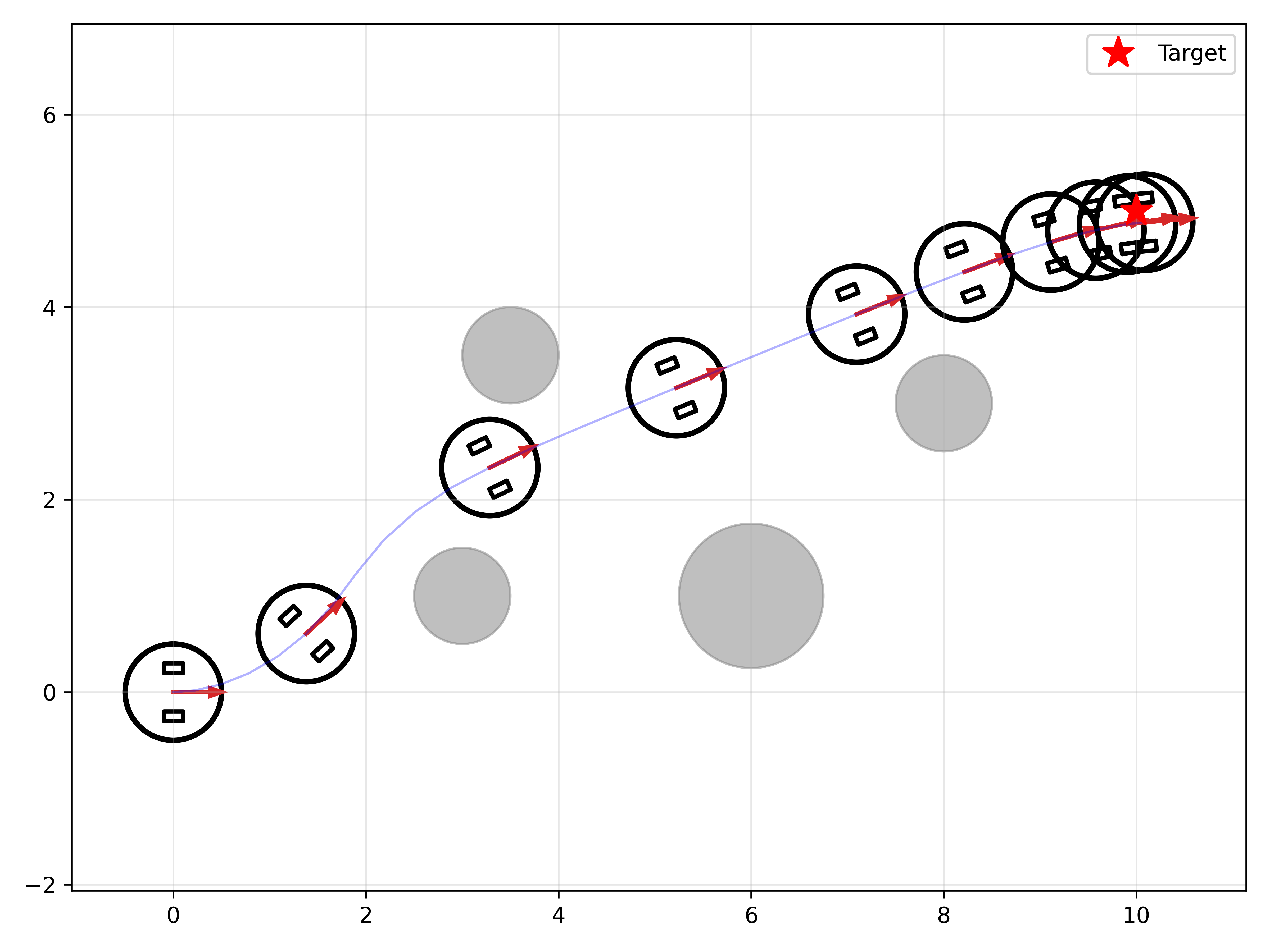}%
    }
    \caption{Test problems. (a) Quad-pendulum $(n_x = 8, n_u = 2)$. The discretization timestep is $\Delta t=\SI{0.02}{\second}$ and the horizon length is $N=200$. 
    (b) Unicycle $(n_x = 4, n_u = 2)$. The discretization timestep is $\Delta t=\SI{0.1}{\second}$ and the horizon length is $N=50$. }
    \label{fig:testing_problems}
\end{figure}

To assess the constraint-handling capability of our solver \textsc{iLQRJax}, we consider the trajectory planning problems for two dynamic models: a unicycle \cite{ChenBranchMPC2022} and a quad-pendulum \cite{SinghSQP2022}, as illustrated in Figure \ref{fig:testing_problems}.
For both problems, we enforce input bounds and collision-avoidance constraints. The collision-avoidance constraint requires the Euclidean distance to exceed a prescribed safety threshold.
We benchmark two AL-based optimal control solvers, \textsc{Trajax}~\cite{trajax} and \textsc{MPX}~\cite{AmatucciMPX2026}, alongside \textsc{IPOPT}~\cite{WachterBiegler2006IPOPT}.
\textsc{Trajax} and \textsc{MPX} are also implemented in JAX.
For all solvers, we set the target constraint tolerance $\epsilon_{\text{tol}}$ to $10^{-5}$. 
We generate $50$ problem instances by randomly sampling initial states around the nominal state while the target state is fixed.
All solvers are initialized with a linearly interpolated trajectory between the initial and target states.
Since \textsc{Trajax} and \textsc{MPX} use practical stopping criteria rather than KKT residuals, we consider a problem solved when both the constraint violation and the complementary slackness are below $\epsilon_{\text{tol}}$. 
The parameters in Algorithm~\ref{alg:alg1} are set to $\tau=100$, $\omega^0=1$, $\eta^0=0.1$, $\alpha_\omega=\beta_\omega=1$, $\alpha_{\eta}=0.1$, and $\beta_{\eta}=0.9$.

\begin{figure}[!htbp]\captionsetup[subfigure]{font=scriptsize}
    \centering
    \subfloat[]{\includegraphics[width=0.5\columnwidth]{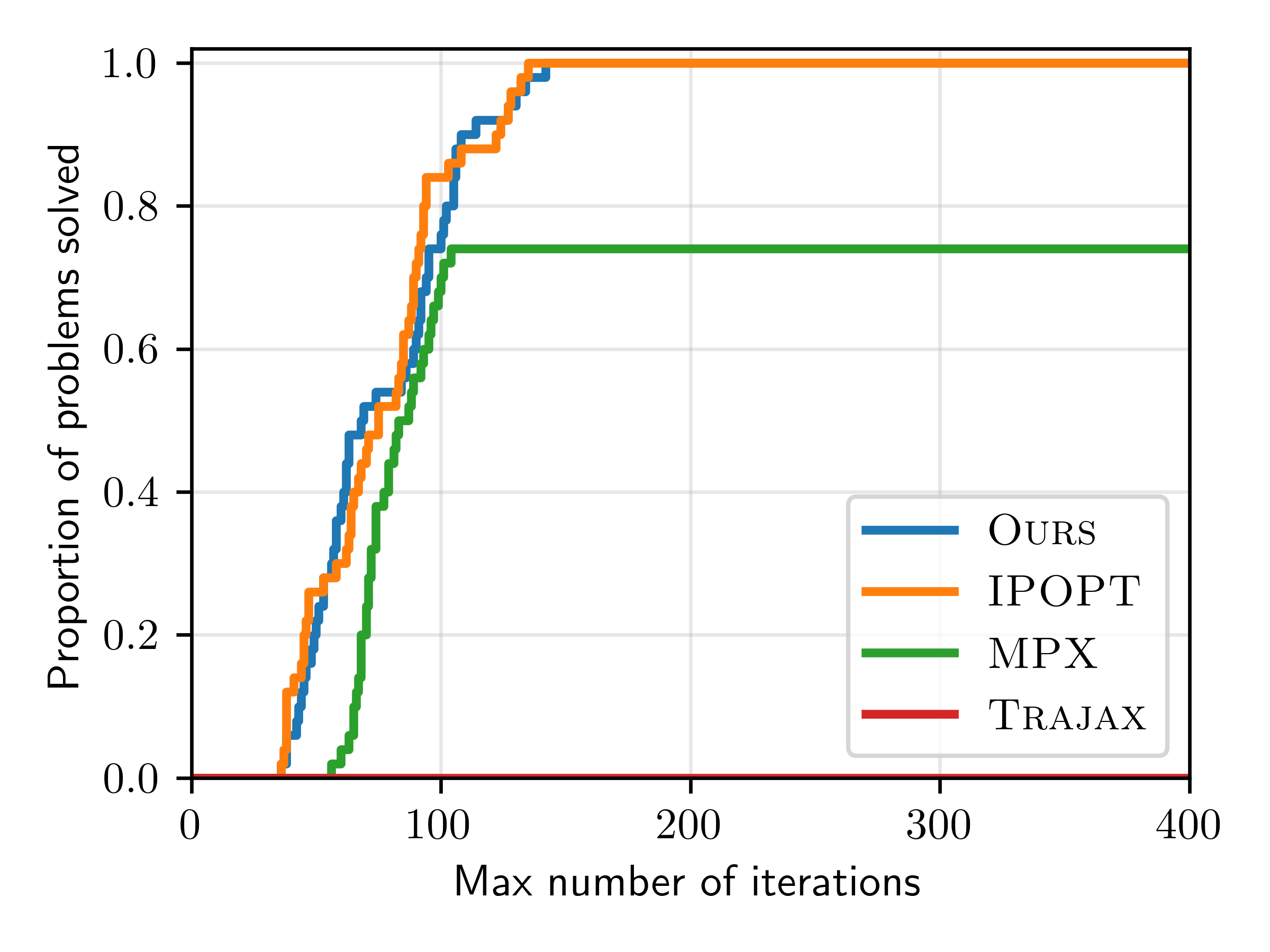}%
    }
    \subfloat[]{\includegraphics[width=0.5\columnwidth]{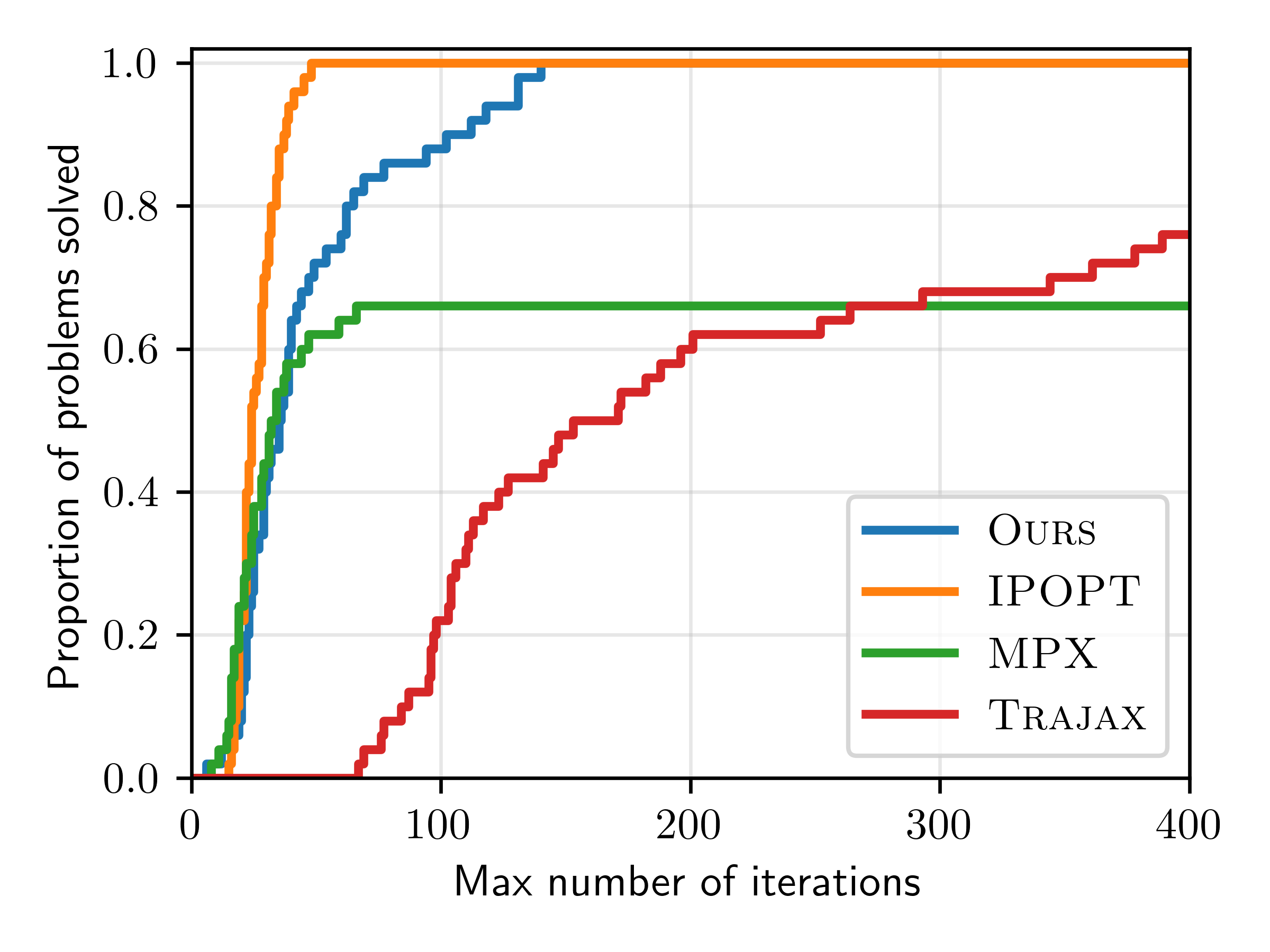}%
    }
    \caption{Proportion of problems solved versus the number of solver iterations. (a) Quad-pendulum. (b) Unicycle.}
    \label{fig:benchmarks}
\end{figure}

Figure~\ref{fig:benchmarks} shows the proportion of problems solved versus the number of iterations. Our solver \textsc{iLQRJax} and \textsc{IPOPT} solve all test problems to the target tolerance, and \textsc{iLQRJax} achieves performance comparable to \textsc{IPOPT} in terms of iteration count. 
Moreover, \textsc{iLQRJax} achieves significantly lower per-iteration computation time than \textsc{IPOPT}. For the unicycle benchmark, the average iteration time is reduced from \SI{3.80}{ms}
to \SI{1.87}{ms}, while for the quad-pendulum benchmark it is reduced from \SI{8.75}{ms} to \SI{4.17}{ms}. These results demonstrate that our solver provides comparable convergence behavior while substantially reducing computational time per iteration.
In contrast, \textsc{Trajax} struggles to solve the quad-pendulum planning problems due to its single-shooting formulation. 
Owing to the multiple-shooting formulation, \textsc{MPX} performs better than \textsc{Trajax}. However, it still fails on the more challenging instances, possibly because its update for the penalty weight associated with the merit function does not follow the theoretical results in \cite{GillAugLag1992} and lacks an adaptive update scheme in the outer loop.

\subsection{Discussion}
The numerical results highlight that selecting a tree LQR solver requires accounting for both the problem scale and the available computational resources.
Although the proposed solver $\textsc{stLQR}$ is highly parallelizable in theory, its higher resource requirements limit its practical scalability to tree optimal control problems with many paths.
Therefore, we also implemented the $\textsc{sLQR}$ solver, which is useful when the performance of $\textsc{stLQR}$ saturates. 
Moreover, \textsc{JAX} relies on the XLA compiler to generate fused kernels and does not explicitly support user-defined kernels.
To further optimize GPU resource utilization, it may be beneficial to implement the parallel algorithm directly in a kernel-based programming language such as CUDA C/C++.
Finally, since most GPUs are optimized for FP32 arithmetic, it may be worthwhile to implement mixed-precision computation or to develop tailored algorithms that improve the numerical stability of FP32.

\FloatBarrier
\section{Conclusion} \label{sec:conclusion}
We have developed a GPU-based optimal control solver tailored for a variety of BMPC problems. 
The tree LQR solver supports different levels of parallelism, enabling it to handle problems at different scales.
Our solver showcases exceptional performance on large-scale problems and can handle general stage-wise constraints in challenging trajectory planning tasks.
One limitation of the current method is that the parallel scan–based solver has high GPU resource demands, which can hinder computational performance.
In the future, we plan to implement the method in CUDA C++ to fully harness the computational power of GPUs.

\begingroup

\appendices
\section{Conditional Value Function} \label{app:dual_value_function}

The conditional value function is the optimal cost of the following fixed-endpoint LQR problem:
\begin{equation} \label{eq:con_value_fun} 
\begin{split} 
     \min_{\boldsymbol{u}_k := (u_n)_{n \in [k, i-1]}} &\sum_{n=k}^{i-1}\ell_{n}(x_{n},u_{n}) := V_{k\to i}(x_{k},x_{i}) \\
    \text{s.t.} &\; x_{n+1} = A_nx_n + B_nu_n + c_n,\\
    &\; \forall n \in [k, i-1].
\end{split} 
\end{equation}
To obtain a condensed formulation of~\eqref{eq:con_value_fun}, we use the dynamics to express the intermediate states as affine functions of the initial state $x_k$ and the control sequence $\boldsymbol{u}_k$. Throughout this derivation, the interval $[k,i]$ is fixed. The condensed vectors and matrices depend on both $k$ and $i$; their dependence on $i$ is suppressed in the notation for readability.
The optimization problem is then
\begin{equation} \label{eq:con_val_fun_dense}
\begin{aligned}
    &V_{k\to i}(x_k, x_i)\\
    &\quad=\min_{\boldsymbol{u}_k}\;
    \begin{bmatrix}
        g_{q,k} \\  g_{r,k} 
    \end{bmatrix} ^\top 
    \begin{bmatrix} x_k \\ \boldsymbol{u}_k \end{bmatrix} {}+\gamma_{k,i}\\
    &\qquad{}+\frac{1}{2}
    \begin{bmatrix} x_k \\ \boldsymbol{u}_k \end{bmatrix} ^\top 
    \begin{bmatrix}
        H_{Q,k} &  H_{M,k}^\top\\ H_{M,k} & H_{R,k}
    \end{bmatrix} 
    \begin{bmatrix} x_k \\ \boldsymbol{u}_k \end{bmatrix}
    \\ 
    &\quad\text{s.t.}\;x_i = \bar{A}_k x_k + \bar{B}_k\boldsymbol{u}_k + \bar{c}_k.
\end{aligned}
\end{equation}
The vector equality constraint enforces the prescribed terminal state $x_i$.
The scalar $\gamma_{k,i}$ collects the constant terms introduced by eliminating the intermediate states. It is independent of the endpoint states and the control sequence and must be retained to preserve the value function. 
The linear coefficients $g_{q,k}$ and $g_{r,k}$, the Hessian blocks $H_{Q,k}$, $H_{M,k}$, and $H_{R,k}$, and the condensed dynamics coefficients $\bar{A}_k$, $\bar{B}_k$, and $\bar{c}_k$ are obtained by eliminating the intermediate states; see~\cite{FrisonCondensing2016} for details of condensing. 

Introducing a multiplier $\lambda\in\mathbb{R}^{n_x}$ for the terminal-state equality constraint gives the min-max formulation of~\eqref{eq:con_val_fun_dense}:
\begin{equation}
\begin{aligned}
    &V_{k\to i}(x_k, x_i)\\
    &\quad=\min_{\boldsymbol{u}_k}\max_{\lambda}\;
    \begin{bmatrix}
        g_{q,k} \\  g_{r,k} 
    \end{bmatrix} ^\top 
    \begin{bmatrix} x_k \\ \boldsymbol{u}_k \end{bmatrix} {}+\gamma_{k,i}\\
    &\qquad{}+\frac{1}{2}
    \begin{bmatrix} x_k \\ \boldsymbol{u}_k \end{bmatrix} ^\top 
    \begin{bmatrix}
        H_{Q,k} &  H_{M,k}^\top\\ H_{M,k} & H_{R,k}
    \end{bmatrix} 
    \begin{bmatrix} x_k \\ \boldsymbol{u}_k \end{bmatrix}
    \\ 
    &\qquad + \lambda^\top \left( x_i - \bar{A}_k x_k - \bar{B}_k\boldsymbol{u}_k - \bar{c}_k \right).
\end{aligned}
\end{equation}
Applying the duality theorem, we interchange the order of the max and min operators, resulting in
\begin{equation*}
    \begin{aligned}
        V_{k\to i}(x_k, x_i) = \max_{\lambda}\; g_{k\rightarrow i}(\lambda;x_k,x_i),
    \end{aligned}
\end{equation*}
where the inner optimization is expressed as
\begin{equation}
    \begin{aligned}
        &g_{k\rightarrow i}(\lambda;x_k,x_i)\\
        &\quad=\min_{\boldsymbol{u}_k}\;\Bigl[\frac{1}{2}\boldsymbol{u}_k^\top H_{R,k}\boldsymbol{u}_k\\
        &\qquad-\left(\lambda^\top\bar{B}_k-g_{r,k}^\top-x_k^\top H_{M,k}^\top\right)\boldsymbol{u}_k\\
        &\qquad+g_{q,k}^\top x_k+\frac{1}{2}x_k^\top H_{Q,k}x_k\\
        &\qquad+\lambda^\top\left(x_i-\bar{A}_kx_k-\bar{c}_k\right)+\gamma_{k,i}\Bigr]\\
        &\quad=\min_{\boldsymbol{u}_k}\;\mathcal{E}(\lambda,\boldsymbol{u}_k,x_k,x_i).
    \end{aligned}
\end{equation}
For each fixed $\lambda$, setting $\nabla_{\boldsymbol{u}_k} \mathcal{E} = 0$ gives the minimizer of the inner problem:
\begin{equation}
    \boldsymbol{u}_k^* = H_{R,k}^{-1} \left( \bar{B}_k^\top \lambda  - g_{r,k} -  H_{M,k} x_k \right) = H_{R,k}^{-1} \xi_k. \label{eq:control_condensed}
\end{equation}
Substituting \eqref{eq:control_condensed} into $\mathcal{E}( \lambda, \boldsymbol{u}_k, x_k, x_i )$ gives
\begin{equation}
    \begin{aligned}
    &g_{k\rightarrow i}(\lambda;x_k,x_i)\\
    &=\gamma_{k,i}-\frac{1}{2}\xi_k^\top H_{R,k}^{-1}\xi_k
      +\lambda^\top(x_i-\bar{A}_kx_k-\bar{c}_k)\\
    &\quad+g_{q,k}^\top x_k+\frac{1}{2}x_k^\top H_{Q,k}x_k\\
    &=\gamma_{k,i}+\frac{1}{2}x_k^\top\left(H_{Q,k}-H_{M,k}^\top H_{R,k}^{-1}H_{M,k}\right)x_k\\
    &\quad+\left(g_{q,k}^\top-g_{r,k}^\top H_{R,k}^{-1}H_{M,k}\right)x_k\\
    &\quad-\frac{1}{2}\lambda^\top\bar{B}_kH_{R,k}^{-1}\bar{B}_k^\top\lambda
      -\frac{1}{2}g_{r,k}^\top H_{R,k}^{-1}g_{r,k}\\
    &\quad+\lambda^\top\Bigl[x_i-\left(\bar{A}_k-\bar{B}_kH_{R,k}^{-1}H_{M,k}\right)x_k\\
    &\hspace{6em}-\left(\bar{c}_k-\bar{B}_kH_{R,k}^{-1}g_{r,k}\right)\Bigr].
    \end{aligned}
\end{equation}
Matching the coefficients in this expression with~\eqref{eq:sol_min} gives
\begin{equation*}
    \begin{aligned}
        P_{k,i} &= H_{Q,k}-H_{M,k}^{\top}H_{R,k}^{-1}H_{M,k},\\
        p_{k,i} &= g_{q,k}-H_{M,k}^{\top}H_{R,k}^{-1}g_{r,k},\\
        C_{k,i} &= \bar{B}_kH_{R,k}^{-1}\bar{B}_k^{\top},\\
        A_{k,i} &= \bar{A}_k-\bar{B}_kH_{R,k}^{-1}H_{M,k},\\
        c_{k,i} &= \bar{c}_k-\bar{B}_kH_{R,k}^{-1}g_{r,k}.
    \end{aligned}
\end{equation*}

\section{Combination Rule} \label{app:combination_rule}
We express the combination of the two conditional value functions, $V_{k\to j}(x_k, x_j)$ and $V_{j\to i}(x_j, x_i)$, as
\begin{equation}\label{eq:min-max}
\begin{aligned}
    &V_{k\to i}(x_k,x_i)\\
    &=\min_{x_j}\Bigl\{\max_{\lambda_1}g_{k\to j}(\lambda_1;x_k,x_j)\\
    &\hspace{4em}+\max_{\lambda_2}g_{j\to i}(\lambda_2;x_j,x_i)\Bigr\}\\
    &=\max_{\lambda_1,\lambda_2}\,\min_{x_j}\Bigl\{g_{k\to j}(\lambda_1;x_k,x_j)\\
    &\hspace{6em}+g_{j\to i}(\lambda_2;x_j,x_i)\Bigr\}\\
    &=z+\max_{\lambda_1,\lambda_2}\,\min_{x_j}\Bigl\{
      \frac{1}{2}x_k^\top P_{k,j}x_k+p_{k,j}^\top x_k\\
    &\quad-\frac{1}{2}\lambda_1^\top C_{k,j}\lambda_1
      +\lambda_1^\top(x_j-A_{k,j}x_k-c_{k,j})\\
    &\quad+\frac{1}{2}x_j^\top P_{j,i}x_j+p_{j,i}^\top x_j\\
    &\quad-\frac{1}{2}\lambda_2^\top C_{j,i}\lambda_2
      +\lambda_2^\top(x_i-A_{j,i}x_j-c_{j,i})\Bigr\}\\
    &=z+\max_{\lambda_1,\lambda_2}\,\min_{x_j}\,
      \Omega(x_j,\lambda_1;x_k,x_i,\lambda_2).
\end{aligned}
\end{equation}
where $z$ is a constant. 
We derive the solution of the min-max problem in \eqref{eq:min-max} over $x_j$ and $\lambda_1$ by setting the gradients of $\Omega$ with respect to the two variables to zero:
\begin{align}
    P_{j,i} x_j &= -\lambda_1 - p_{j,i} + A^\top_{j,i}\lambda_2, \label{eq:grad_x} \\
    x_j &= C_{k,j}\lambda_1 + A_{k,j}x_k + c_{k,j}. \label{eq:grad_lambda}
\end{align}
Combining \eqref{eq:grad_x} and \eqref{eq:grad_lambda}, we obtain the expression for $\lambda_1$:
\begin{equation}
\begin{aligned}
    \lambda_1&=(I+P_{j,i}C_{k,j})^{-1}\Bigl[-p_{j,i}+A^\top_{j,i}\lambda_2\\
    &\hspace{8em}-P_{j,i}(A_{k,j}x_k+c_{k,j})\Bigr].
\end{aligned}\label{eq:lambda1}
\end{equation}
Substituting \eqref{eq:grad_lambda} into $\Omega$ in \eqref{eq:min-max}, we obtain
\begin{align*}
    \Omega &= 
    \frac{1}{2} c_{k,j}^\top p_{j,i} + 
    \frac{1}{2}x_{k}^{\top}P_{k,j}x_{k} + \left( p_{k,j} + \frac{1}{2}A_{k,j}^\top p_{j,i} \right)^{\top} x_{k} \\
    &\quad -\frac{1}{2}\lambda_2 ^{\top }C_{j,i}\lambda_2
      +\lambda_2^\top\Bigl(x_i-\frac{1}{2}A_{j,i}A_{k,j}x_k\\
    &\hspace{9em}-c_{j,i}-\frac{1}{2}A_{j,i}c_{k,j}\Bigr)\\
    &\quad + \frac{1}{2} \Big(  
    \lambda_1^\top C_{k,j} p_{j,i} 
    - \lambda_1^\top C_{k,j} A^\top_{j,i} \lambda_2
    \\[-0.5ex]
    &\hspace{3em}{}-x_k^\top A_{k,j}^\top\lambda_1-c_{k,j}^\top\lambda_1\Big).
\end{align*}
Next, we eliminate $\lambda_1$ using \eqref{eq:lambda1}. The resulting expression includes quadratic terms in \( x_k \) and \( \lambda_2 \), 
linear terms in \( x_k \) and \( \lambda_2 \), as well as a cross term involving \( x_k \) and \( \lambda_2 \). 
By matching these terms with those in \eqref{eq:sol_min}, we obtain
\begin{equation}    
\begin{aligned} \label{eq:p_ki}
    P_{k,i} &= P_{k,j} + A_{k,j}^\top (I + P_{j,i}C_{k,j})^{-1} P_{j,i} A_{k,j}, \\
    p_{k,i} &= p_{k,j} + \frac{1}{2}\Bigl[-A_{k,j}^\top P_{j,i}(I+C_{k,j}P_{j,i})^{-1}C_{k,j}p_{j,i}\\
    &\quad+A_{k,j}^\top p_{j,i}\\
    &\quad+A_{k,j}^\top P_{j,i}(I+C_{k,j}P_{j,i})^{-1}c_{k,j}\\
    &\quad+A_{k,j}^\top(I+P_{j,i}C_{k,j})^{-1}(P_{j,i}c_{k,j}+p_{j,i})\Bigr]
\end{aligned} 
\end{equation}
We can further simplify the expression for $p_{k,i}$ in \eqref{eq:p_ki} using the Woodbury matrix identity, yielding
\begin{align*}
    p_{k,i}
    &= p_{k,j} + \frac{1}{2}\left[A_{k,j}^\top \left( I - P_{j,i} \left( I + C_{k,j}P_{j,i} \right)^{-1} C_{k,j} \right) p_{j,i}  \right. \\
    &\quad \left. + A_{k,j}^\top \left(I + P_{j,i}C_{k,j} \right)^{-1} P_{j,i} c_{k,j} \right. \\
    &\quad \left. + A_{k,j}^\top (I + P_{j,i}C_{k,j})^{-1} \left( P_{j,i}c_{k,j} + p_{j,i} \right) \right] \\
    &= p_{k,j} + \frac{1}{2}\left[A_{k,j}^\top \left(I + P_{j,i}C_{k,j} \right)^{-1} p_{j,i}  \right. \\
    &\quad \left. + A_{k,j}^\top \left(I + P_{j,i}C_{k,j} \right)^{-1} P_{j,i} c_{k,j} \right. \\
    &\quad \left. + A_{k,j}^\top (I + P_{j,i}C_{k,j})^{-1} \left( P_{j,i}c_{k,j} + p_{j,i} \right)
    \right] \\
    &= p_{k,j} + A_{k,j}^\top (I + P_{j,i}C_{k,j})^{-1} \left( P_{j,i}c_{k,j} + p_{j,i} \right)
\end{align*}
Similarly, we can derive $C_{k,i}$, $A_{k,i}$ and $c_{k,i}$ as follows:
\begin{align*}
    C_{k,i} &= C_{j,i} + A_{j,i} \left( I + C_{k,j}P_{j,i} \right)^{-1} C_{k,j} A^\top_{j,i}, \\
     A_{k,i} &= -\frac{1}{2}\left[ A_{j,i} C_{k,j} \left( I + P_{j,i}C_{k,j} \right)^{-1} P_{j,i} A_{k,j} - A_{j,i}A_{k,j} \right. \\ 
    &\quad \quad \quad - \left. A_{j,i} \left( I + C_{k,j}P_{j,i} \right)^{-1} A_{k,j}  
    \right] \\
    &= A_{j,i} \left( I + C_{k,j}P_{j,i} \right)^{-1} A_{k,j}, \\
    c_{k,i} &= c_{j,i} - \frac{1}{2} \left[ A_{j,i} \left( I + C_{k,j}P_{j,i} \right)^{-1} \left( C_{k,j} p_{j,i} - c_{k,j}  \right) \right. \\
    & \left. + A_{j,i} C_{k,j} \left( I + P_{j,i}C_{k,j} \right)^{-1}p_{j,i} \right. \\
    & \left. - A_{j,i}c_{k,j} + A_{j,i} C_{k,j} \left( I + P_{j,i}C_{k,j} \right)^{-1} P_{j,i}c_{k,j} \right] \\
    &= c_{j,i} + A_{j,i} \left( I + C_{k,j}P_{j,i} \right)^{-1} \left( c_{k,j} - C_{k,j} p_{j,i}  \right).
\end{align*}
\section{Chain Value Function}
\label{app:chain_value_recovery}

This appendix derives~\eqref{eq:chain_condition} and~\eqref{eq:chain_inlet_value} from the chain minimization problem~\eqref{eq:V_s}.
Using the dual representation of the chain CVF in~\eqref{eq:sol_min}, we obtain
\begin{equation}\label{eq:chain_saddle}
\begin{aligned}
&V_s(x_s)=\min_{x_m}\max_{\lambda}\;\Phi(x_s,x_m,\lambda),\\
&\Phi(x_s,x_m,\lambda)\\
&\quad=z+\frac12 x_s^\top P_{s,m}x_s+p_{s,m}^\top x_s
  -\frac12\lambda^\top C_{s,m}\lambda\\
&\qquad+\lambda^\top(x_m-A_{s,m}x_s-c_{s,m})\\
&\qquad+\frac12 x_m^\top P_mx_m+p_m^\top x_m,
\end{aligned}
\end{equation}
where $z$ collects terms independent of $x_s$, $x_m$, and $\lambda$.
Under the convex quadratic assumptions, the saddle point satisfies the first-order conditions
\begin{equation}\label{eq:chain_stationarity}
\begin{aligned}
\nabla_{x_m}\Phi&=\lambda^*+P_mx_m^*+p_m=0,\\
\nabla_{\lambda}\Phi&=x_m^*-A_{s,m}x_s-c_{s,m}-C_{s,m}\lambda^*=0.
\end{aligned}
\end{equation}
Substituting $\lambda^*=-P_mx_m^*-p_m$ into the second equation in~\eqref{eq:chain_stationarity} gives
\begin{equation}\label{eq:chain_terminal_system}
\underbrace{(I+C_{s,m}P_m)}_{E_{s,m}}x_m^*
=A_{s,m}x_s+c_{s,m}-C_{s,m}p_m.
\end{equation}
The matrix $E_{s,m}$ is nonsingular when $C_{s,m}$ and $P_m$ are symmetric positive semidefinite, as discussed after~\eqref{eq:comb_rule}.
Writing $x_m^*=\tilde A_{s,m}x_s+\tilde c_{s,m}$ and matching the coefficients of $x_s$ and the constant terms in~\eqref{eq:chain_terminal_system} leads to~\eqref{eq:chain_condition}.

To recover the coefficients of $V_s$, we differentiate the optimal value with respect to $x_s$.
Since $x_m^*$ and $\lambda^*$ depend on $x_s$, we write
$V_s(x_s)=\Phi\bigl(x_s,x_m^*(x_s),\lambda^*(x_s)\bigr)$.
The chain rule gives
\begin{equation*}
\begin{aligned}
\nabla_{x_s}V_s(x_s)
&=\nabla_{x_s}\Phi
  +\left(\frac{\partial x_m^*}{\partial x_s}\right)^\top\nabla_{x_m}\Phi\\
&\quad+\left(\frac{\partial\lambda^*}{\partial x_s}\right)^\top\nabla_{\lambda}\Phi,
\end{aligned}
\end{equation*}
where all gradients of $\Phi$ are evaluated at $(x_s,x_m^*,\lambda^*)$.
The first-order conditions~\eqref{eq:chain_stationarity} give $\nabla_{x_m}\Phi=0$ and $\nabla_{\lambda}\Phi=0$, so the second and third terms are zero, and only $\nabla_{x_s}\Phi$ remains.
Using~\eqref{eq:chain_saddle} and substituting $\lambda^*=-P_mx_m^*-p_m$ and $x_m^*=\tilde A_{s,m}x_s+\tilde c_{s,m}$, we obtain
\begin{equation}\label{eq:chain_value_gradient}
\begin{aligned}
\nabla_{x_s}V_s(x_s)
&=P_{s,m}x_s+p_{s,m}-A_{s,m}^\top\lambda^*\\
&=P_{s,m}x_s+p_{s,m}+A_{s,m}^\top(P_mx_m^*+p_m)\\
&=\bigl(P_{s,m}+A_{s,m}^\top P_m\tilde A_{s,m}\bigr)x_s\\
&\quad+p_{s,m}+A_{s,m}^\top(p_m+P_m\tilde c_{s,m}).
\end{aligned}
\end{equation}
Comparing this expression with $\nabla_{x_s}V_s(x_s)=P_sx_s+p_s$ yields
\begin{equation*}
\begin{aligned}
P_s&=P_{s,m}+A_{s,m}^\top P_m\tilde A_{s,m},\\
p_s&=p_{s,m}+A_{s,m}^\top(p_m+P_m\tilde c_{s,m}),
\end{aligned}
\end{equation*}
which is~\eqref{eq:chain_inlet_value}.
\par
\endgroup

\printbibliography

@ARTICLE{WensingOptimization2024,
  author={Wensing, Patrick M. and Posa, Michael and Hu, Yue and Escande, Adrien and Mansard, Nicolas and Prete, Andrea Del},
  journal={IEEE Transactions on Robotics}, 
  title={Optimization-Based Control for Dynamic Legged Robots}, 
  year={2024},
  volume={40},
  number={},
  pages={43-63},
  doi={10.1109/TRO.2023.3324580}}

@ARTICLE{ChenCiLQR2019,
  author={Chen, Jianyu and Zhan, Wei and Tomizuka, Masayoshi},
  journal={IEEE Transactions on Intelligent Vehicles}, 
  title={Autonomous Driving Motion Planning With Constrained Iterative {LQR}}, 
  year={2019},
  volume={4},
  number={2},
  pages={244-254},
  doi={10.1109/TIV.2019.2904385}}

@INPROCEEDINGS{NguyenAerial2021,
  author={Nguyen, Huan and Kamel, Mina and Alexis, Kostas and Siegwart, Roland},
  booktitle={2021 European Control Conference (ECC)}, 
  title={Model Predictive Control for Micro Aerial Vehicles: A Survey}, 
  year={2021},
  volume={},
  number={},
  pages={1556-1563},
  doi={10.23919/ECC54610.2021.9654841}}

@inproceedings{Blelloch1990PrefixSA,
  title={Prefix sums and their applications},
  author={Guy E. Blelloch},
  year={1990},
  url={https://api.semanticscholar.org/CorpusID:60459178}
}

@ARTICLE{SarkkaTemporal2023,
  author={Särkkä, Simo and García-Fernández, Ángel F.},
  journal={IEEE Transactions on Automatic Control}, 
  title={Temporal Parallelization of Dynamic Programming and Linear Quadratic Control}, 
  year={2023},
  volume={68},
  number={2},
  pages={851-866},
  doi={10.1109/TAC.2022.3147017}}

@INPROCEEDINGS{FrisonCondensing2016,
  author={Frison, Gianluca and Kouzoupis, Dimitris and Jørgensen, John Bagterp and Diehl, Moritz},
  booktitle={2016 IEEE 55th Conference on Decision and Control (CDC)}, 
  title={An efficient implementation of partial condensing for Nonlinear Model Predictive Control}, 
  year={2016},
  volume={},
  number={},
  pages={4457-4462},
  doi={10.1109/CDC.2016.7798946}}

@article{FRISON201714399,
title = {A high-performance {Riccati} based solver for tree-structured quadratic programs},
journal = {IFAC-PapersOnLine},
volume = {50},
number = {1},
pages = {14399-14405},
year = {2017},
note = {20th IFAC World Congress},
issn = {2405-8963},
doi = {https://doi.org/10.1016/j.ifacol.2017.08.2027},
url = {https://www.sciencedirect.com/science/article/pii/S2405896317326666},
author = {Gianluca Frison and Dimitris Kouzoupis and Moritz Diehl and John Bagterp Jørgensen},
}

@ARTICLE{ChenBranchMPC2022,
  author={Chen, Yuxiao and Rosolia, Ugo and Ubellacker, Wyatt and Csomay-Shanklin, Noel and Ames, Aaron D.},
  journal={IEEE Robotics and Automation Letters}, 
  title={Interactive Multi-Modal Motion Planning With Branch Model Predictive Control}, 
  year={2022},
  volume={7},
  number={2},
  pages={5365-5372},
  doi={10.1109/LRA.2022.3156648}}

@INPROCEEDINGS{AlsterdaContingency2019,
  author={Alsterda, John P. and Brown, Matthew and Gerdes, J. Christian},
  booktitle={2019 American Control Conference (ACC)}, 
  title={Contingency Model Predictive Control for Automated Vehicles}, 
  year={2019},
  volume={},
  number={},
  pages={717-722},
  doi={10.23919/ACC.2019.8815260}}

@ARTICLE{HardyContingencyPlanning2013,
  author={Hardy, Jason and Campbell, Mark},
  journal={IEEE Transactions on Robotics}, 
  title={Contingency Planning Over Probabilistic Obstacle Predictions for Autonomous Road Vehicles}, 
  year={2013},
  volume={29},
  number={4},
  pages={913-929},
  doi={10.1109/TRO.2013.2254033}}

@ARTICLE{BatkovicRobustSMPC2021,
  author={Batkovic, Ivo and Rosolia, Ugo and Zanon, Mario and Falcone, Paolo},
  journal={IEEE Control Systems Letters}, 
  title={A Robust Scenario {MPC} Approach for Uncertain Multi-Modal Obstacles}, 
  year={2021},
  volume={5},
  number={3},
  pages={947-952},
  doi={10.1109/LCSYS.2020.3006819}}

@article{kouzoupis_dual_2019,
	title = {A dual {Newton} strategy for tree‐sparse quadratic programs and its implementation in the open‐source software {treeQP}},
	volume = {29},
	issn = {1049-8923, 1099-1239},
	url = {https://onlinelibrary.wiley.com/doi/10.1002/rnc.4503},
	doi = {10.1002/rnc.4503},
	number = {8},
	journal = {International Journal of Robust and Nonlinear Control},
	author = {Kouzoupis, D. and Klintberg, E. and Frison, G. and Gros, S. and Diehl, M.},
	year = {2019},
	pages = {2438--2457},
}

@inproceedings{sinha_real-time_2024,
  author = {Rohan Sinha and Amine Elhafsi and Christopher Agia and Matthew Foutter and Edward Schmerling and Marco Pavone},
  title     = {Real-Time Anomaly Detection and Reactive Planning with Large Language Models},
  booktitle = {Proceedings of Robotics: Science and Systems},
  address   = {Delft, Netherlands},
  month     = jul,
  year      = {2024},
}

@article{wright_partitioned_1991,
	title = {Partitioned Dynamic Programming for Optimal Control},
	volume = {1},
	issn = {1052-6234},
	url = {https://epubs.siam.org/doi/10.1137/0801037},
	doi = {10.1137/0801037},
	number = {4},
	urldate = {2024-06-03},
	journal = {SIAM Journal on Optimization},
	author = {Wright, Stephen J.},
	month = nov,
	year = {1991},
	note = {Publisher: Society for Industrial and Applied Mathematics},
	pages = {620--642},
}

@inproceedings{nielsen_parallel_2015,
	address = {Osaka},
	title = {A parallel structure exploiting factorization algorithm with applications to Model Predictive Control},
	doi = {10.1109/CDC.2015.7402830},
	booktitle = {2015 54th {IEEE} {Conference} on {Decision} and {Control} ({CDC})},
	publisher = {IEEE},
	author = {Nielsen, Isak and Axehill, Daniel},
	month = dec,
	year = {2015},
	pages = {3932--3938},
}

@inproceedings{jallet_parallel_2024,
	title = {Parallel and Proximal Linear-Quadratic Methods for Real-Time Constrained Model-Predictive Control},
	author = {Jallet, Wilson and Dantec, Ewen and Arlaud, Etienne and Carpentier, Justin and Mansard, Nicolas},
	booktitle = {Proceedings of Robotics: Science and Systems},
    address   = {Delft, Netherlands},
    month     = jul,
    year      = {2024},
}

@ARTICLE{JalletProxDDP2025,
  author={Jallet, Wilson and Bambade, Antoine and Arlaud, Etienne and El-Kazdadi, Sarah and Mansard, Nicolas and Carpentier, Justin},
  journal={IEEE Transactions on Robotics}, 
  title={{ProxDDP}: Proximal Constrained Trajectory Optimization}, 
  year={2025},
  volume={41},
  number={},
  pages={2605-2624},
  doi={10.1109/TRO.2025.3554437}}

@article{BettsLowThrust2003,
author = {Betts, John T. and Erb, Sven O.},
title = {Optimal Low Thrust Trajectories to the Moon},
journal = {SIAM Journal on Applied Dynamical Systems},
volume = {2},
number = {2},
pages = {144-170},
year = {2003},
doi = {10.1137/S1111111102409080},

URL = { 
    https://doi.org/10.1137/S1111111102409080
},
eprint = { 
    https://doi.org/10.1137/S1111111102409080
},
}

@INPROCEEDINGS{HeTDROBCA2021,
  author={He, Runxin and Zhou, Jinyun and Jiang, Shu and Wang, Yu and Tao, Jiaming and Song, Shiyu and Hu, Jiangtao and Miao, Jinghao and Luo, Qi},
  booktitle={2021 American Control Conference (ACC)}, 
  title={{TDR-OBCA}: A Reliable Planner for Autonomous Driving in Free-Space Environment}, 
  year={2021},
  volume={},
  number={},
  pages={2927-2934},
  doi={10.23919/ACC50511.2021.9483020}}

@inproceedings{laine_parallelizing_2019,
	address = {Nice, France},
	title = {Parallelizing {LQR} Computation Through Endpoint-Explicit {Riccati} Recursion},
	doi = {10.1109/CDC40024.2019.9029974},
	booktitle = {2019 {IEEE} 58th {Conference} on {Decision} and {Control} ({CDC})},
	publisher = {IEEE},
	author = {Laine, Forrest and Tomlin, Claire},
	month = dec,
	year = {2019},
	pages = {1395--1402},
}

@article{marti_improving_2015,
	title = {Improving scenario decomposition algorithms for robust nonlinear model predictive control},
	volume = {79},
	issn = {0098-1354},
	url = {https://www.sciencedirect.com/science/article/pii/S0098135415001258},
	doi = {10.1016/j.compchemeng.2015.04.024},
	urldate = {2024-09-23},
	journal = {Computers \& Chemical Engineering},
	author = {Martí, Rubén and Lucia, Sergio and Sarabia, Daniel and Paulen, Radoslav and Engell, Sebastian and de Prada, César},
	month = aug,
	year = {2015},
	pages = {30--45},
}

@article{sampathirao_massively_2024,
	title = {Massively parallelizable proximal algorithms for large-scale stochastic optimal control problems},
	volume = {45},
	issn = {1099-1514},
	doi = {10.1002/oca.3054},
	number = {1},
	urldate = {2024-05-26},
	journal = {Optimal Control Applications and Methods},
	author = {Sampathirao, Ajay K. and Patrinos, Panagiotis and Bemporad, Alberto and Sopasakis, Pantelis},
	year = {2024},
	pages = {45--63},
}

@incollection{Plancher_performance_2020,
	address = {Cham},
	title = {A Performance Analysis of Parallel Differential Dynamic Programming on a {GPU}},
	volume = {14},
	urldate = {2024-04-30},
	booktitle = {Algorithmic {Foundations} of {Robotics} {XIII}},
	publisher = {Springer International Publishing},
	author = {Plancher, Brian and Kuindersma, Scott},
	year = {2020},
	pages = {656--672},
}

@INPROCEEDINGS{FarshidianMPC2017,
  author={Farshidian, Farbod and Jelavic, Edo and Satapathy, Asutosh and Giftthaler, Markus and Buchli, Jonas},
  booktitle={2017 IEEE-RAS 17th International Conference on Humanoid Robotics (Humanoids)}, 
  title={Real-time motion planning of legged robots: A model predictive control approach}, 
  year={2017},
  volume={},
  number={},
  pages={577-584},
  doi={10.1109/HUMANOIDS.2017.8246930}}

@article{DengParallel2019,
title = {A parallel {Newton}-type method for nonlinear model predictive control},
journal = {Automatica},
volume = {109},
pages = {108560},
year = {2019},
issn = {0005-1098},
doi = {https://doi.org/10.1016/j.automatica.2019.108560},
url = {https://www.sciencedirect.com/science/article/pii/S0005109819304212},
author = {Haoyang Deng and Toshiyuki Ohtsuka},
}

@INPROCEEDINGS{TassaiLQR2012,
  author={Tassa, Yuval and Erez, Tom and Todorov, Emanuel},
  booktitle={2012 IEEE/RSJ International Conference on Intelligent Robots and Systems}, 
  title={Synthesis and stabilization of complex behaviors through online trajectory optimization}, 
  year={2012},
  volume={},
  number={},
  pages={4906-4913},
  doi={10.1109/IROS.2012.6386025}}

@INPROCEEDINGS{HowellALTRO2019,
  author={Howell, Taylor A. and Jackson, Brian E. and Manchester, Zachary},
  booktitle={2019 IEEE/RSJ International Conference on Intelligent Robots and Systems (IROS)}, 
  title={ALTRO: A Fast Solver for Constrained Trajectory Optimization}, 
  year={2019},
  volume={},
  number={},
  pages={7674-7679},
  doi={10.1109/IROS40897.2019.8967788}}

@INPROCEEDINGS{LiUnified2023,
  author={Li, He and Yu, Wenhao and Zhang, Tingnan and Wensing, Patrick M.},
  booktitle={2023 IEEE/RSJ International Conference on Intelligent Robots and Systems (IROS)}, 
  title={A Unified Perspective on Multiple Shooting In Differential Dynamic Programming}, 
  year={2023},
  volume={},
  number={},
  pages={9978-9985},
  doi={10.1109/IROS55552.2023.10342217}}

@book{Nocedal_numerical_2006,
        author = {Nocedal, Jorge and Wright, Stephen J.},
	title = {Numerical {Optimization}},
	publisher = {Springer New York},
	year = {2006},
	doi = {10.1007/978-0-387-40065-5},
}

@article{schraudolph_fast_2002,
	title = {Fast Curvature Matrix-Vector Products for Second-Order Gradient Descent},
	volume = {14},
	doi = {10.1162/08997660260028683},
	number = {7},
	journal = {Neural Computation},
	author = {Schraudolph, Nicol N.},
	month = jul,
	year = {2002},
	pages = {1723--1738},
}

@ARTICLE{GrandiaLocomotion2023,
  author={Grandia, Ruben and Jenelten, Fabian and Yang, Shaohui and Farshidian, Farbod and Hutter, Marco},
  journal={IEEE Transactions on Robotics}, 
  title={Perceptive Locomotion Through Nonlinear Model-Predictive Control}, 
  year={2023},
  volume={39},
  number={5},
  pages={3402-3421},
  doi={10.1109/TRO.2023.3275384}}

@software{trajax,
  author = {Google},
  month = {8},
  title = {Trajax: A Python library for differentiable optimal control on accelerators.},
  url = {https://github.com/google/trajax},
  year = {2021}
}

@ARTICLE{WangNonGaussian2020,
  author={Wang, Allen and Jasour, Ashkan and Williams, Brian C.},
  journal={IEEE Robotics and Automation Letters}, 
  title={Non-{Gaussian} Chance-Constrained Trajectory Planning for Autonomous Vehicles Under Agent Uncertainty}, 
  year={2020},
  volume={5},
  number={4},
  pages={6041-6048},
  doi={10.1109/LRA.2020.3010755}}

@ARTICLE{AhnSafeMotionPlanning2022,
  author={Ahn, Heejin and Chen, Colin and Mitchell, Ian M. and Kamgarpour, Maryam},
  journal={IEEE Control Systems Letters}, 
  title={Safe Motion Planning Against Multimodal Distributions Based on a Scenario Approach}, 
  year={2022},
  volume={6},
  number={},
  pages={1142-1147},
  doi={10.1109/LCSYS.2021.3089641}}

@article{zhang_automated_2024,
title = {Automated Lane Merging via Game Theory and Branch Model Predictive Control},
author = {Zhang, Luyao and Han, Shaohang and Grammatico, Sergio},
journal = {IEEE Transactions on Control Systems Technology},
year = {2024},
}

@software{jax2018github,
  author = {James Bradbury and Roy Frostig and Peter Hawkins and Matthew James Johnson and Chris Leary and Dougal Maclaurin and George Necula and Adam Paszke and Jake Vander{P}las and Skye Wanderman-{M}ilne and Qiao Zhang},
  title = {{JAX}: composable transformations of {P}ython+{N}um{P}y programs},
  url = {http://github.com/jax-ml/jax},
  version = {0.3.13},
  year = {2018},
}

@article{ConnBCL1991,
author = {Conn, Andrew R. and Gould, Nicholas I. M. and Toint, Philippe},
title = {A Globally Convergent Augmented {Lagrangian} Algorithm for Optimization with General Constraints and Simple Bounds},
journal = {SIAM Journal on Numerical Analysis},
volume = {28},
number = {2},
pages = {545-572},
year = {1991},
doi = {10.1137/0728030},
}

@INPROCEEDINGS{PhiquepalControlTree2021,
  author={Phiquepal, Camille and Toussaint, Marc},
  booktitle={2021 IEEE International Conference on Robotics and Automation (ICRA)}, 
  title={Control-Tree Optimization: an approach to {MPC} under discrete Partial Observability}, 
  year={2021},
  volume={},
  number={},
  pages={9666-9672},
  doi={10.1109/ICRA48506.2021.9561664}}

@InProceedings{GillSQP2012,
    author="Gill, Philip E. and Wong, Elizabeth",
    editor="Lee, Jon and Leyffer, Sven",
    title="Sequential Quadratic Programming Methods",
    booktitle="Mixed Integer Nonlinear Programming",
    year="2012",
    publisher="Springer New York",
    address="New York, NY",
    pages="147--224",
}

@article{GillSNOPT2002,
author = {Gill, Philip E. and Murray, Walter and Saunders, Michael A.},
title = {SNOPT: An {SQP} Algorithm for Large-Scale Constrained Optimization},
journal = {SIAM Journal on Optimization},
volume = {12},
number = {4},
pages = {979-1006},
year = {2002},
doi = {10.1137/S1052623499350013},

URL = { 
        https://doi.org/10.1137/S1052623499350013
}
}

@incollection{GillAugLag1992,
  author    = {Gill, Philip E. and Murray, Walter and Saunders, Michael A. and Wright, Margaret H.},
  title     = {Some Theoretical Properties of an Augmented {Lagrangian} Merit Function},
  booktitle = {Advances in Optimization and Parallel Computing},
  editor    = {Pardalos, P. M.},
  publisher = {North-Holland},
  address   = {Amsterdam},
  year      = {1992},
  pages     = {101--128}
}

@INPROCEEDINGS{SinghSQP2022,
  author={Singh, Sumeet and Slotine, Jean-Jacques and Sindhwani, Vikas},
  booktitle={2022 International Conference on Robotics and Automation (ICRA)}, 
  title={Optimizing Trajectories with Closed-Loop Dynamic {SQP}}, 
  year={2022},
  volume={},
  number={},
  pages={5249-5254},
  doi={10.1109/ICRA46639.2022.9811562}}

@article{WachterBiegler2006IPOPT,
  author    = {W{\"a}chter, Andreas and Biegler, Lorenz T.},
  title     = {On the Implementation of an Interior-Point Filter Line-Search Algorithm for Large-Scale Nonlinear Programming},
  journal   = {Mathematical Programming},
  volume    = {106},
  number    = {1},
  pages     = {25--57},
  year      = {2006},
  doi       = {10.1007/s10107-004-0559-y}
}

@ARTICLE{AmatucciMPX2026,
  author={Amatucci, Lorenzo and Sousa-Pinto, João and Turrisi, Giulio and Orban, Dominique and Barasuol, Victor and Semini, Claudio},
  journal={IEEE Robotics and Automation Letters}, 
  title={Primal-Dual {iLQR} for {GPU}-Accelerated Learning and Control in Legged Robots}, 
  year={2026},
  volume={11},
  number={1},
  pages={1010-1017},
  doi={10.1109/LRA.2025.3632610}}

@article{FRISON2020HPIPM,
title = {{HPIPM}: a high-performance quadratic programming framework for model predictive control},
journal = {IFAC-PapersOnLine},
volume = {53},
number = {2},
pages = {6563-6569},
year = {2020},
note = {21st IFAC World Congress},
issn = {2405-8963},
doi = {https://doi.org/10.1016/j.ifacol.2020.12.073},
url = {https://www.sciencedirect.com/science/article/pii/S2405896320303293},
author = {Gianluca Frison and Moritz Diehl},
}

@article{rockafellar1976augmented,
  author    = {R. T. Rockafellar},
  title     = {Augmented {Lagrangians} and Applications of the Proximal Point Algorithm in Convex Programming},
  journal   = {Mathematics of Operations Research},
  volume    = {1},
  number    = {2},
  pages     = {97--116},
  year      = {1976},
  publisher = {INFORMS},
  doi       = {10.1287/moor.1.2.97}
}

@article{LuciaSMPC2013,
title = {Multi-stage nonlinear model predictive control applied to a semi-batch polymerization reactor under uncertainty},
journal = {Journal of Process Control},
volume = {23},
number = {9},
pages = {1306-1319},
year = {2013},
issn = {0959-1524},
doi = {https://doi.org/10.1016/j.jprocont.2013.08.008},
url = {https://www.sciencedirect.com/science/article/pii/S0959152413001686},
author = {Sergio Lucia and Tiago Finkler and Sebastian Engell},
}

\end{document}